\documentclass[12pt]{article}
\usepackage{graphicx}
\usepackage{amsfonts}
\usepackage{amsmath}
\usepackage[utf8]{inputenc}
\usepackage[english]{babel}
\usepackage{amssymb,amsmath,amsthm}
\usepackage{url}

\newtheorem{theorem}{Theorem}
\DeclareUnicodeCharacter{2061}{}

\usepackage{times}

\newenvironment{sciabstract}{%
\begin{quote} \bf}
{\end{quote}}

\title{Finding the Best Route  During the Pandemic Disease} 

\author
{Amirsadegh Mirgalooyebayat,$^{1\ast}$ Farzad Didehvar$^{1\ast}$\\
\\
\normalsize{$^{1}$Department of Mathematics and Computer Science, Amir Kabir University of Technology}
\\
\normalsize{$^\ast$E-mail: amir.bayat@aut.ac.ir,  didehvar@aut.ac.ir}
}

\date{February 2024}

\begin{document} 

\baselineskip24pt

\maketitle

\begin{sciabstract}
\section*{Abstract}
  In this article, we try to find the best routes during the pandemic so that the probability of contracting the disease is the lowest. According to the results of this article, we can design software to find the best route.
\\
\\
Kew Words:
\\
Covid-19, pandemic, mathematical model, routing applications, routing.  
\end{sciabstract}

\section*{Introduction}
After the 2019-2021 Covid-19 pandemic, many questions about pandemics and outbreaks have been revived. Before that and in 2013,
Huppert, Amit
and Katriel, Guy
 in a review article investigated the application of mathematical modeling in predicting and understanding the spread of infectious diseases. The authors demonstrated the application of mathematical models in infectious disease epidemics through various examples. They explained how the models help to estimate key parameters such as the initial reproduction number
\footnote{$R_0$}
,which represents the average number of infections caused by an infected person. The models also help to understand the impact of interventions such as vaccination, social distancing and quarantine measures.\cite{huppert2013mathematical}

One of the issues that was taken into consideration in these years was estimating how to transmit Covid-19 and calculating the probability of getting the disease. In 2020 Lelieveld et al. in an article, evaluated the risk of Covid-19 infection through aerosol particles in indoor environments. This paper uses mathematical modeling and simulation to estimate how SARS-CoV-2, the virus that causes covid-19, is transmitted
through the air. In this study, the role of aerosol particles, which can remain suspended in the air for a long time, is emphasized as a factor of transmission of Covid-19. Also, a model is presented to calculate the concentration of viral airborne particles over time and space in an indoor environment. Using these calculations, they estimated the risk of infection based on various factors such as room size, number of occupants, duration of exposure, and ventilation rate.\cite{lelieveld2020model}
Next, in 2021, Guzman, Marcelo I. has a review on the effect of the size of the carrier components of the coronavirus RNAs. Also, the person carrying the virus can spread the virus components within a distance of 2 meters by talking, sneezing or coughing, and these components remain stagnant in the ambient air for up to 3 hours.
\cite{guzman2021overview}

Another issue was how the virus spreads through the air and the importance of ventilation. In 2021, Li, Yuguo et al. investigated a case in which airborne transmission of the SARS-CoV-2 virus occurred in a poorly ventilated restaurant. By analyzing an outbreak among customers, this study emphasizes the potential for airborne transmission of the virus in enclosed spaces. In 2020, using numerical simulation tools for modeling, Vuorinen, Ville et al. studied the transmission of aerosol particles in indoor environments and assessed the risks associated with inhalation of these particles carrying SARS-CoV-2. In this study, the risk of virus exposure through aerosols in different indoor conditions, including the effects of changes in ventilation, room crowding and different activity levels, is calculated.\cite{vuorinen2020modelling} In 2020, Anderson, Elizabeth L et al. investigated the importance of aerosol transmission in the spread of Covid-19 and its implications for public health measures. They examined factors that affect aerosol distribution in indoor environments, such as airflow patterns, humidity, and temperature.
\cite{anderson2020consideration}

Another thing that was noticed in these years is the effect of various environmental factors on disease transmission. In 2020, Feng, Yu et al. used numerical simulations to investigate how environmental factors such as wind and relative humidity might affect the effectiveness of social distancing as a measure to prevent airborne transmission of Covid-19.\cite{feng2020influence} In 2021, Hassan, Md Nazmul et al. presented a mathematical model to analyze and predict the spread of Covid-19 in Texas with the aim of investigating the possibility of disease outbreak and providing insight for effective control measures. The study used historical data on Covid-19 cases, hospitalizations and deaths in Texas to calibrate the mathematical model. Various factors such as population density, mobility patterns and social distancing measures were included in the model to capture the dynamics of the epidemic.\cite{hassan2023mathematical}

In 2020, Miralles-Pechu{\'a}n, Luis et al. investigated the impact of quarantine and travel restrictions on public health. Using the SEIR model and with the help of two Deep Q-learning and genetics algorithms, they presented a model that allowed governments to make better decisions in different phases of the Covid-19 epidemic.\cite{miralles2020methodology}

One of the things to consider is the possibility of catching the disease and its approximate duration in passenger cars and public transportation. In 2022, Sarhan, Abd Alhamid R et al. investigated different scenarios to understand how the virus would spread in the confined space of a passenger car. This study adjusted various conditions such as ventilation patterns, presence and use of masks, passenger position and travel duration to model the spread of the virus. In this article, estimates have been made about the possibility of passengers being infected in different conditions, including the identification of people at risk depending on their seats in the car and the speed of the car's ventilation.\cite{sarhan2022numerical} In 2024, Zhao, Yu et al. investigated the characteristics of the dynamic patterns of droplets in city buses using numerical simulations. The results showed that in 30 to 90 seconds the droplets begin to spread throughout the entire compartment and affect almost all passengers, and the droplets sprayed by the standing passengers spread faster and within 30 seconds affected 8 passengers around the target passenger.\cite{zhao2024characteristics} 

Another important issue in these years is the impact of the pandemic on public transportation. In 2020, Gkiotsalitis, Konstantinos and Cats, Oded reviewed the factors affecting public transportation during a pandemic. The authors examine how public transportation systems have adapted to changing transportation patterns, health and safety requirements, and economic pressures resulting from the public health crisis.\cite{gkiotsalitis2021public} In 2022, Marra, Alessio D et al. investigated the effects of the Covid-19 pandemic on the use of public transportation and route selection. This study shows that during the pandemic, people show more aversion to crowded spaces and are more concerned about maintaining social distance. As a result, they prioritize routes with less traffic and shorter travel times to minimize exposure to other passengers.\cite{marra2022impact}

The problem that has been considered in this article is to find the routes within the city in such a way that it has the lowest probability of contracting a pandemic disease. In 2020, Chen, Dawei et al. developed a vehicle routing solution that enables contactless delivery while optimizing the operational efficiency of the distribution network.\cite{chen2020vehicle} In 2020, Pacheco, Joaqu{\'\i}n and Laguna, Manuel presented an innovative method to find the optimal way to manufacture and deliver face shields during the Covid-19 pandemic. Their research was conducted in the city of Burgos in Spain, and the goal was to optimize the delivery time of raw materials to small face shield manufacturing companies and then deliver the manufactured shields to health centers and hospitals.\cite{pacheco2020vehicle} In 2021, using the backtracking algorithm and the adapted traveling salesman problem, P{\u{a}}curar, Cristina Maria et al. developed a creative method to find the optimal route for tourists during the Covid-19 pandemic.\cite{puacurar2021tourist} In 2021, Eren, Emre and Tuzkaya, Umut R{\i}fat investigated vehicle routing specifically in the context of medical waste collection during the Covid-19 pandemic. The authors' goal was to find a vehicle routing solution that would optimize the collection process while also taking into account safe distance requirements during a pandemic.\cite{eren2021safe}

In our opinion, things can be progressed with the mathematical modeling that is done in this article, and the main goal of the article is to present an algorithm based on the mathematical model to find a path with the least probability of getting the disease.

In the second chapter, we seek to find a mathematical relationship for the probability of getting the disease in different ways. It is assumed that each route consists of sub-routes traveled by one of the forms of "walking", "subway", "BRT", "bus" or "car". In this chapter, a mathematical model for the probability of getting the disease is first presented, and then we calculate the coefficients of this model for each of the types of sub-paths. In the following, with the assumption that each route is a combination of these sub-routes, a relationship for the probability of getting the disease in the combination of routes is presented, which can be used to calculate the probability of getting the disease in all types of routes.

In the third chapter, we apply the model defined in the second chapter to different routes between a specific origin and destination and determine the best route according to the lowest probability of getting the disease.
For this purpose and to determine the route, the routing applications "Nashan" and "Balad" were used for the routes between Sadeghieh Square and Amirkabir University in Tehran at 18:00 on a mid-week day.
In this chapter, the routing algorithm is written in general mode. Then, the number of affected people and the average distance between people in different routes have been calculated in general. Then, a general relationship for the probability of getting the disease in each of the possible environments has been rewritten according to the relationships obtained in the previous chapter. Finally, the relationship of getting the disease in different environments has been determined in a specific state and the probability of getting the disease for the different paths suggested in the "Nashan" and "Balad" applications has been calculated.

In the fourth chapter, the summary and conclusion have been done. The results indicate that the probability of getting the disease in each route can be calculated by knowing the suggested routes between the origin and specific destination. For example, in this article, the suggested routes of "Nashan" and "Balad" applications between "Sadeghieh Square" and "Amirkabir University" at 18:00 on a mid-week day were examined and the best route in each application was determined. At the end, suggestions for future work were presented.

\section*{Finding a Relationship to the Probability of contracting a Disease}
In this section, we seek to find a mathematical relationship for the probability of getting the disease in different ways. It is assumed that each route consists of sub-routes traveled by one of the forms of "walking", "subway", "BRT", "bus" or "car". In an environment where two people are present and one of them is suffering from the COVID-19 disease, the probability of catching the disease in the other person can be calculated. The dimensions of the environment, the time that people are in the environment and the activity level of people (equivalent to air inhalation in liters per hour) are used as input values to calculate the probability of getting the disease.\footnote{https://covid-19.forhealth.org/covid-19-transmission-calculator/}
Other parameters such as the amount of ventilation or the opening and closing of the windows are also influential in these calculations, which we have omitted from their changes with simplifying assumptions. Having this probability for different environments and specifying the mathematical model, the coefficients of this model are calculated for Covid-19. In the following, a relationship for the probability of getting the disease in the combination of these paths is also presented.

\subsection*{Combination of Multiple Paths}
\begin{figure}[h!]
		\centering 
		\includegraphics[width=\linewidth]{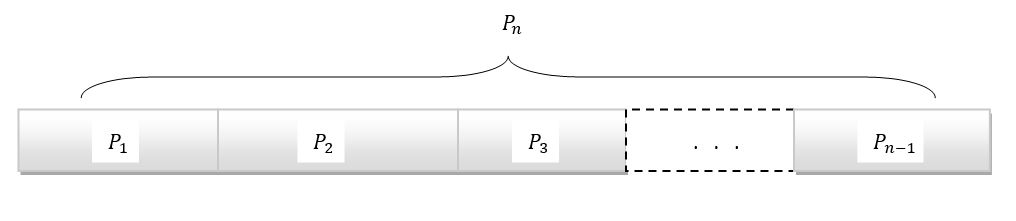}
		\caption{Combined route}
		\label{fig:fig1_1}
\end{figure}

Each route can be divided into sub-routes so that the final route is a combination of these sub-routes. In this section, we seek to find a relationship between the probability of getting the disease in combined routes. The question that arises is that if we have the probability of getting the disease in the paths  $1,2,3,...,n-1$, what is the probability of getting the disease in the path $n$ which is the combination of these paths along each other? The combined route is generally specified in the figure $\eqref{fig:fig1_1}$.

\begin{theorem}
If path $n$ consists of $n-1$   paths along each other and all these paths are independent of each other and the probability of getting the disease in the $i$th path is equal to $P_i$, the probability of getting the disease in the $n$th path is equal to :
\begin{equation}
	P_n = 1-\prod_{i=1}^{n-1} (1- P_i)
	\label{eq0}
\end{equation}
\end{theorem}

\begin{proof}
The probability of not getting sick during the path $i$ is equal to $1-P_i$. Since the paths are two by two independent of each other, the probability of not getting sick at the end of the path $n$ (that is, the probability of not getting sick after passing through all the paths) is equal to the multiplicative combination of the probability of not getting sick in each path. That is, this probability is equal to:
\begin{align*}
	\prod_{i=1}^{n-1} (1-P_i)
\end{align*}	
So, the probability of getting sick at the end of the journey is equal to the complement of the above probability and is equal to:
\begin{align*}
	1-\prod_{i=1}^{n-1} (1- P_i)
\end{align*}
\end{proof}

\subsection*{Relationship Between Probability of Contracting Disease and Path Parameters}
Assuming three key independent parameters influence the probability of contracting a disease, the probability of contracting a disease on an arbitrary path will be a function f(x,y,z) of these three parameters. These three parameters are environmental density (x), the probability that a person is a carrier (y), and the duration the person spends in the environment (z).

The meaning of the environmental density is the number of people per unit area under consideration. Also, the probability that a person is a carrier is a parameter that should be determined experimentally or from articles.

Next, we will examine the time parameter and its relationship with the probability of getting the disease.

\subsection*{Duration in the Environment}
One of the factors influencing the transmission of the disease is the length of time the person under investigation is exposed to the disease. In this section, we seek to find a relationship between the probability of getting the disease according the time spent in the environment. For this purpose, we first state and prove some theorems.

\begin{theorem}
If a person is in the environment during the time $a+b$, the probability of getting the disease in terms of time is obtained from the following relationship:
\begin{equation}
	f(a+b)=f(a)+f(b)-f(a)f(b)
	\label{eq22}	
\end{equation}
\end{theorem}

\begin{proof}
As in the case where a path was a combination of several paths, here also if the time that a person is in the environment increases from $z$ to $2z$, it is like in the third path that is a combination of two paths with the same time $ z$ is placed.
   Also, if a person is in the environment for the duration of $a+b$, it is the same as having traveled a combination of paths for the duration of $a$ and $b$. So according to the relation $\eqref{eq0}$ we will have:
  \begin{align*}
  	f(a+b)=1-(1-f(a))(1-f(b))=1-1+f(a)+f(b)-f(a)f(b) \\
  	\Rightarrow f(a+b)=f(a)+f(b)-f(a)f(b)
  \end{align*}
\end{proof}

\begin{theorem}
If the duration of time a person is in the environment is equal to $z$ and $k$ is a natural and constant number, for all arbitrary and natural numbers $n$, the probability of getting a disease in terms of time can be calculated from the following relationship:
\begin{equation}
	f(kf^{-1} (\frac{1}{n}))=1-(\frac{n-1}{n})^k ;k,n \in \mathbb {N}
\end{equation}
\end{theorem}

\begin{proof}
In relation $\eqref{eq22}$, if we put $a=z$ and $b=f^{-1}({\frac{1}{n}})$ so that $n$ is a natural number, we will have:
\begin{equation}
	f(z+b)=f(z)+\frac{1}{n}-\frac{1}{n} f(z)=\frac{n-1}{n} f(z)+\frac{1}{n}
	\label{eq23}
\end{equation}
Now we prove the problem by induction.

The base of induction:
\begin{align*}
	k=1 \Rightarrow f(f^{-1}\frac{1}{n})=\frac{1}{n}=1-(\frac{n-1}{n})
\end{align*}
So the basis of induction is established.

Induction step:
To prove the induction, we put in the relation $\eqref{eq23}$:
 $z=(k-1)b$
. will have:
\begin{align*}
	f(kb)=\frac{n-1}{n} f((k-1)b)+\frac{1}{n}
\end{align*}
By placing the assumption of induction in the above relation, we have:
\begin{align*}
	f(kb)=\frac{n-1}{n} (1-(\frac{n-1}{n})^{(k-1)})+\frac{1}{n}
	=1-(\frac{n-1}{n})^k
\end{align*}
Since $b=f^{-1}({\frac{1}{n}})$ then the induction theorem is proven.
\end{proof}

Now we look for an explicit relation for $f(z)$:
\begin{displaymath}
	f(a)=\frac{1}{n} \Rightarrow f(ka)=1-(1-\frac{1}{n})^k=1-(1-f(a))^k
\end{displaymath}
Now, if we have $z=ka$, then we reach the following relationship:
\begin{equation}
	f(z)=1-(1-f(\frac{z}{k}))^k
	\label{eq24}
\end{equation}

\begin{theorem}
The probability of getting the disease in terms of time is obtained from the following relationship:
\begin{equation}
	f(z)=1-e^{cz}
\end{equation}
\end{theorem}

\begin{proof}
Since in relation $\eqref{eq24}$, the function $f(z)$ is independent of $k$, for two different values of $k$, the following equality holds:
\begin{equation}
	(1-f(\frac{z}{k_1} ))^(k_1 )=(1-f(\frac{z}{k_2} ))^(k_2 )
\end{equation}
As a result, we will have:
\begin{displaymath}
	k_1  log⁡(1-f(\frac{z}{k_1} ))=k_2  log⁡(1-f(\frac{z}{k_2} )) \Rightarrow\\
	 \frac{k_1}{k_2} =\frac{log⁡(1-f(\frac{z}{k_2} ))}{log⁡(1-f(\frac{z}{k_1} ))} 
\end{displaymath}
From the above relationship, we can conclude that the function $f(z)$ is as follows:
\begin{align*}
	f(z)=1-e^{cz}
\end{align*}
\end{proof}

Finally, the function $f(x,y,z)$ becomes as follows:
\begin{equation}
	f(x,y,z)=1-e^{c(x,y)z}
	\label{eq9}
\end{equation}
Next, the coefficient $c(x,y)$ should be calculated.

\subsection*{Modeling Disease Rate}
In this section, we design a page with dimensions $m \times l$ and randomly place n infected people in the cells. An uninfected person passes through this path and at the end, we calculate the probability of them contracting the disease.

We assume the time duration that the person spends in cell j is $T_j$. Also, the distance of the person when they are in cell j to the infected person i is $r_{ij}$. Since the probability of contracting the disease has an inverse relationship with distance from the infected person and a direct relationship with the duration in the environment, the probability of contracting the disease in cell j is obtained from the following relationship:
\begin{equation}	
	g(\sum_{i=1}^n \frac{k}{r_{ji}^2})T_j
	\label{eq2}	
\end{equation}

With the reasonable assumption that the probability of contracting the disease in different cells is independent, the probability that the person does not contract the disease throughout the path is the multiplicative combination of not contracting the disease in each of the cells:
\begin{displaymath}
	\prod_{j=1}^n {1-g(\sum_{i=1}^n \frac{k}{r_{ji}^2})T_j}
\end{displaymath}
Where A is the path traversed by the uninfected person. Finally, the probability of contracting the disease at the end of the path equals:
\begin{equation}
	h(k)= 1-\prod_{j=1}^m {1-g(\sum_{i=1}^n \frac{k}{r_{ji}^2})T_j}
\end{equation}

\begin{figure}[h!]
		\centering 
		\includegraphics[width=\linewidth]{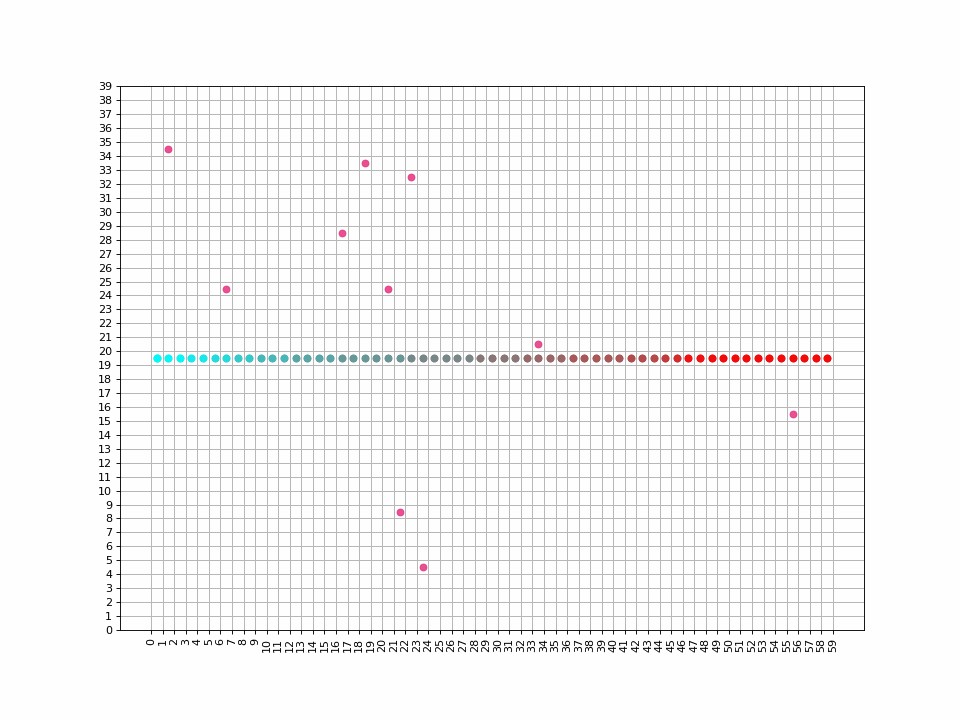}
		\caption{Passing a person without a disease through a $60 \times 40$ screen with 10 sick people present}
		\label{fig:fig1_2}
\end{figure}

\subsection*{Calculating Model Coefficients}
The inputs of the problem are considered as follows:
Input 1: Total population (s), environmental density (x), the probability that a person is a carrier (y), and duration the person spends in the environment (z).

Note: For simplicity, we assume a person either is a carrier or not, and we do not consider the degree of being a carrier as a probability function. With this assumption, we arrive at the following relationship:
\begin{equation}
	y=\frac{n}{s}
\end{equation}
As a result, instead of Input 1, we can use Input 2.

Input 2: Carrier population (n), environmental density (x), duration person spends in the environment (z)

Also, the output of the problem is defined as follows:

Output: Probability of contracting disease f(x,y,z)

To calculate disease contraction probability, we computed this value in two ways and it is enough to equate their results:
\begin{equation}
	f(x,y,z)=h(k) \Rightarrow 1-e^{c(x,y)z} = 1-\prod_{j=1}^m {1-g(\sum_{i=1}^n \frac{k}{r_{ji}^2})T_j}
\end{equation}

We guess that we need to write the initial relationship such that we ultimately arrive at the following relationship:
\begin{equation}
	c(x,y)=\frac{1}{z} \sum_{j=1}^m{P(ln(1-\sum_{i=1}^n{\frac{kz}{mr_{ij}^2}}))}; P(t)=1-e^t
	\label{eq1}
\end{equation}

By substituting P(t) in equation \eqref{eq1}, we arrive at the following relationship:
\begin{equation}
	c(x,y)=\frac{1}{z}\sum_{j=1}^m{\sum_{i=1}^n{\frac{kz}{mr_{ij}^2}}}=
	\sum_{j=1}^m{\sum_{i=1}^n{\frac{k}{mr_{ij}^2}}}
	\label{eq5}
\end{equation}

In this case, the value c(x,y) is independent of time.

By substituting the value of c(x,y) in equation \eqref{eq9}, we arrive at the following relationship which gives the probability of contracting the disease:
\begin{equation}
	f(x,y,z)=1-e^{\sum_{j=1}^m{\sum_{i=1}^n{\frac{kz}{mr_{ij}^2}}}}
	\label{eq10}
\end{equation}

Now to obtain k, we proceed as follows:

In an experiment, two people are placed in a room for a specific duration and at a certain distance. We assume the length and width of the room is 1 unit and there is only one infected person in the room. That is, in equation \eqref{eq5} we set n = 1 and m = 1. Thus, we arrive at the following relationships:
\begin{equation}
	c(x,y)=\frac{k}{r^2}
\end{equation}

\begin{equation}
	f(x,y,z)=1-e^{\frac{kz}{r^2}}
	\label{eq6}
\end{equation}

Now if we have the value of the probability of contracting the disease f(x,y,z), we can calculate k as follows:
\begin{equation}
	\frac{kz}{r^2}=ln(1-f(x,y,z)) \Rightarrow k=\frac{r^2}{z}ln(1-f(x,y,z))
	\label{eq7}
\end{equation}

Also, if there are two infected people in the environment instead of one, the above relationships become:
\begin{equation}
	c(x,y)=\frac{k}{r_1^2}+\frac{k}{r_2^2}=k(\frac{1}{r_1^2}+\frac{1}{r_2^2})
\end{equation}

\begin{equation}
	f(x,y,z)=1-e^{kz(\frac{1}{r_1^2}+\frac{1}{r_2^2})} \Rightarrow
	kz(\frac{1}{r_1^2}+\frac{1}{r_2^2})=ln(1-f(x,y,z)) \Rightarrow
	k=\frac{r_1^2r_2^2}{z(r_1^2+r_2^2)}ln(1-f(x,y,z))
\end{equation}

Now we need to look in research papers for the probability of contracting COVID between two people over a specified duration and use that to calculate the appropriate coefficient k.

\begin{figure}[h!]
		\centering 
		\includegraphics[width=\linewidth]{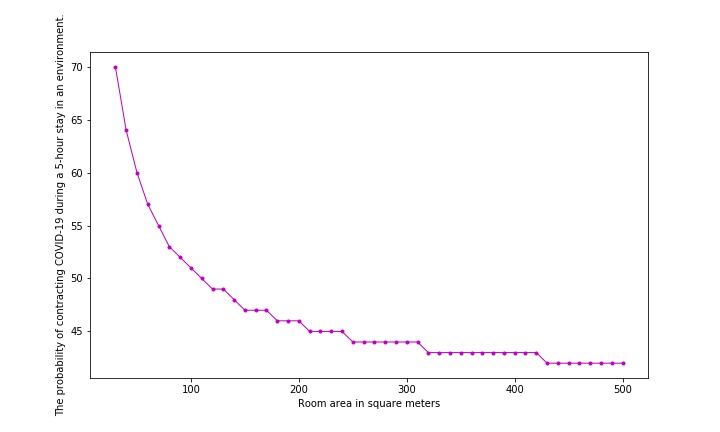}
		\caption{The probability of catching COVID-19 in 5 hours of being in the environment according to the size of the room for low activity}
		\label{fig:fig1_0}
\end{figure}

For this purpose, we have used the information available in reference \cite{covid-19} Below, we have calculated the probability of contracting COVID in rooms with different areas and durations. It is assumed that one carrier is present in the room and the average distance between two people is 6 feet. The results were that as the area increases to a large number like 500 square meters, the probability of COVID also converges to a specific number, an example of which is plotted in Figure \eqref{fig:fig1_0}

Thus, it can be concluded that the probability of contracting COVID in an open space is equivalent to that of a very large room. This calculation is presented in four tables \eqref{tab1}, \eqref{tab2}, \eqref{tab3}, and \eqref{tab4} for four conditions of sitting resting or mild, moderate, or intense activity in an open space. In the first case, this coefficient is a number between -0.053442 and -0.033613, in the second case a number between -0.373153 and -0.352379, in the third case a number between -1.120676- and -1.172772-, and in the fourth case a number between -5.134-012 and -4704.739. This amount is calculated by substituting the assumed values in equation \eqref{eq7}.

\begin{table} [!h]
  \centering
      \caption{Coefficient
     $k$
     in a room with a size of 500 square meters and the activity of sitting or resting for different activity times}
    \label{tab1}
    \begin{tabular}{|c|c|c|}
    \hline
         \textbf{Time in hours} & \textbf{Percentage of infection} & \textbf{$k$} \\
    \hline  
         1 & 1 & $-0.033613$ \\
    \hline 
         2 & 3 & $-0.050936$ \\
    \hline 
         3 & 4 & $-0.045510$ \\
    \hline 
         4 & 6 & $-0.051736$ \\
    \hline 
         5 & 7 & $-0.048543$ \\
    \hline 
         6 & 9 & $-0.052570$ \\
    \hline 
         7 & 10 & $-0.050340$ \\
    \hline 
         8 & 12 & $-0.053442$ \\
    \hline 
         9 & 13 & $-0.051751$ \\
    \hline
         10 & 14 & $-0.050443$ \\
    \hline  
    \end{tabular}
\end{table}

\begin{table} [!h]
  \centering
      \caption{Coefficient
     $k$
     in a room with a size of 500 square meters and low activity for different activity times}
    \label{tab2}
    \begin{tabular}{|c|c|c|}
    \hline
         \textbf{Time in hours} & \textbf{Percentage of infection} & \textbf{$k$} \\
    \hline  
         1 & 10 & $-0.352379$ \\
    \hline 
         2 & 20 & $-0.373153$ \\
    \hline 
         3 & 28 & $-0.366228$ \\
    \hline 
         4 & 35 & $-0.360189$ \\
    \hline 
         5 & 42 & $-0.364369$ \\
    \hline 
         6 & 48 & $-0.364511$ \\
    \hline 
         7 & 53 & $-0.360740$ \\
    \hline 
         8 & 58 & $-0.362670$ \\
    \hline 
         9 & 63 & $-0.369476$ \\
    \hline
         10 & 67 & $-0.370793$ \\
    \hline  
    \end{tabular}
\end{table}

\begin{table} [!h]
  \centering
      \caption{Coefficient
     $k$
     in a room with a size of 500 square meters and moderate activity for different activity times}
    \label{tab3}
    \begin{tabular}{|c|c|c|}
    \hline
         \textbf{Time in hours} & \textbf{Percentage of infection} & \textbf{$k$} \\
    \hline  
         1 & 30 & $-1.192903$ \\
    \hline 
         2 & 51 & $-1.192903$ \\
    \hline 
         3 & 66 & $-1.202696$ \\
    \hline 
         4 & 76 & $-1.193251$ \\
    \hline 
         5 & 83 & $-1.185265$ \\
    \hline 
         6 & 88 & $-1.181874$ \\
    \hline 
         7 & 92 & $-1.206760$ \\
    \hline 
         8 & 94 & $-1.176185$ \\
    \hline 
         9 & 96 & $-1.196173$ \\
    \hline
         10 & 97 & $-1.172772$ \\
    \hline  
    \end{tabular}
\end{table}

\begin{table} [!h]
  \centering
      \caption{Coefficient
     $k$
     in a room with a size of 500 square meters and intense activity for different activity times}
    \label{tab4}
    \begin{tabular}{|c|c|c|}
    \hline
         \textbf{Time in hours} & \textbf{Percentage of infection} & \textbf{$k$} \\
    \hline  
         1 & 76 & $-4.773004$ \\
    \hline 
         2 & 94 & $-4.704739$ \\
    \hline 
         3 & 99 & $-5.134012$ \\
    \hline 
    \end{tabular}
\end{table}

Based on the information obtained, it can be concluded that the coefficient value depends on the level of activity and is approximately constant for a specific activity level. So in equation \eqref{eq6}, instead of a constant coefficient k, we should use the function k(E) which indicates that this coefficient is a function of the energy consumption level in the environment.

Below we calculate the coefficient k(E) in the open environment and derive the relationship for the probability of contracting the disease in this environment.

\subsubsection*{Calculating Coefficient k(E) and Probability of Contracting Disease in Open Environment}
In this section, to calculate coefficient k(E) the activity level is classified into four categories, where the average air intake per hour in liters is specified for each category \cite{united1989exposure}. This information is provided in Table \eqref{tab5}.

\begin{table} [!h]
  \centering
      \caption{Average air inhalation in liters per hour for different activity levels}
    \label{tab5}
    \begin{tabular}{|c|c|c|}
    \hline
         \textbf{Type of activity} & \textbf{amount of air inhaled} \\
    \hline  
    	Sitting and resting & $300$ \\
    \hline  
    Low activity & $780$ \\
    \hline  
    Moderated activity & $1740$ \\
    \hline  
   Intense activity & $3180$ \\
    \hline  
    \end{tabular}
\end{table}

To calculate k(E) the value is merged in four tables \eqref{tab1}, \eqref{tab2}, \eqref{tab3}, and \eqref{tab4} with the numerical activity level given in Table \eqref{tab5}. In Python, the correlation between coefficient k and activity level was calculated to be -0.93076, indicating that these two parameters have a high correlation and that change in activity level affects coefficient k. Also, using the linear regression model in Python, coefficient k(E) was calculated as follows:
\begin{equation}
	k(E) = -0.00143853E+0.71455401
	\label{eq8}
\end{equation}

Where the R-Score in this model is 0.866 indicating good model accuracy.

\begin{figure}[h!]
		\centering 
		\includegraphics[width=\linewidth]{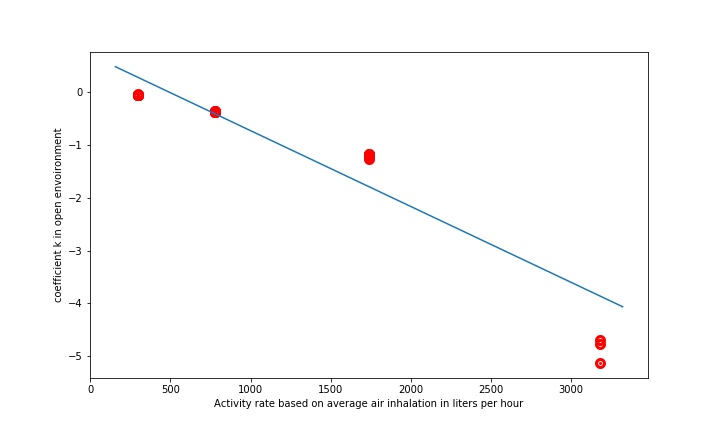}
		\caption{The value of $k(E)$ according to the amount of activity along with the estimated linear regression model in the open environment}
		\label{fig:fig1_3}
\end{figure}

Figure \eqref{fig:fig1_3} shows the relationship between k(E) versus activity level along with the estimated linear regression model.

Finally, by substituting the k(E) value from equation \eqref{eq8} in equation (\eqref{eq10}, we arrive at equation \eqref{eq11} which represents the probability of contracting disease in an open environment.
\begin{equation}
	f(m,n,z,E)=1-e^{\sum_{j=1}^m{\sum_{i=1}^n{\frac{(-0.00143853E+0.71455401)z}{mr_{ij}^2}}}}
	\label{eq11}
\end{equation}

Next, we calculate the k(E) coefficient for other environments.

\subsubsection*{Calculating Coefficient k(E) and Probability of Contracting Disease in Subway Wagons}
The dimensions of a Tehran subway wagon are $1.952 \times 2.6$ meters \cite{MetroDimensions}. We consider this environment as an enclosed space with proper ventilation and a 50 square meters area. The information in tables \eqref{tab6}, \eqref{tab7}, \eqref{tab8} and \eqref{tab9} has been calculated for four conditions of sitting (resting or mild, moderate or intense activity) in subway wagons \cite{covid-19}.
\begin{table} [!h]
  \centering
      \caption{Coefficient
    $k$
  in a room with a size of 50 square meters and the activity of sitting or resting for different activity times}
    \label{tab6}
    \begin{tabular}{|c|c|c|}
    \hline
         \textbf{Time in hours} & \textbf{Percentage of infection} & \textbf{$k$} \\
    \hline  
         1 & 2 & $-0.067568$ \\
    \hline 
         2 & 4 & $-0.068265$ \\
    \hline 
         3 & 7 & $-0.080904$ \\
    \hline 
         4 & 9 & $-0.078856$ \\
    \hline 
         5 & 11 & $-0.077950$ \\
    \hline 
         6 & 13 & $-0.077627$ \\
    \hline 
         7 & 16 & $-0.083304$ \\
    \hline 
         8 & 18 & $-0.082965$ \\
    \hline 
         9 & 20 & $-0.082923$ \\
    \hline
         10 & 22 & $-0.083098$ \\
    \hline  
    \end{tabular}
\end{table}

\begin{table} [!h]
  \centering
      \caption{Coefficient $k$ in a room with 50 square meters and low activity for different activity times}
    \label{tab7}
    \begin{tabular}{|c|c|c|}
    \hline
         \textbf{Time in hours} & \textbf{Percentage of infection} & \textbf{$k$} \\
    \hline  
         1 & 14 & $-0.504429$ \\
    \hline 
         2 & 27 & $-0.526277$ \\
    \hline 
         3 & 37 & $-0.515094$ \\
    \hline 
         4 & 47 & $-0.530839$ \\
    \hline 
         5 & 55 & $-0.534123$ \\
    \hline 
         6 & 62 & $-0.539349$ \\
    \hline 
         7 & 68 & $-0.544407$ \\
    \hline 
         8 & 73 & $-0.547385$ \\
    \hline 
         9 & 77 & $-0.546149$ \\
    \hline
         10 & 81 & $-0.555433$ \\
    \hline  
    \end{tabular}
\end{table}

\begin{table} [!h]
  \centering
      \caption{Coefficient
    $k$
 in a room with a size of 50 square meters and moderated activity for different activity times}
    \label{tab8}
    \begin{tabular}{|c|c|c|}
    \hline
         \textbf{Time in hours} & \textbf{Percentage of infection} & \textbf{$k$} \\
    \hline  
         1 & 39 & $-1.653179$ \\
    \hline 
         2 & 63 & $-1.662643$ \\
    \hline 
         3 & 78 & $-1.688005$ \\
    \hline 
         4 & 87 & $-1.705884$ \\
    \hline 
         5 & 92 & $-1.689465$ \\
    \hline 
         6 & 95 & $-1.669876$ \\
    \hline 
         7 & 97 & $-1.675388$ \\
    \hline 
         8 & 98 & $-1.635475$ \\
    \hline 
         9 & 99 & $-1.711337$ \\
    \hline
         10 & 99 & $-1.540204$ \\
    \hline  
    \end{tabular}
\end{table}

\begin{table} [!h]
  \centering
      \caption{Coefficient
    $k$
  in a room with a size of 50 square meters and intense activity for different activity times}
    \label{tab9}
    \begin{tabular}{|c|c|c|}
    \hline
         \textbf{Time in hours} & \textbf{Percentage of infection} & \textbf{$k$} \\
    \hline  
         1 & 86 & $-6.575683$ \\
    \hline 
         2 & 98 & $-6.541899$ \\
    \hline 
    \end{tabular}
\end{table}

Similar to the open environment case, by comparing the values of the four tables above with the average air intake for different activities given in Table \eqref{tab5}, we have calculated coefficient k(E) for the subway wagon environment in Python. The correlation coefficient between the coefficient and activity level in this case is -0.925089. Also, using the linear regression model in Python, coefficient k(E) was calculated as follows:
\begin{equation}
	k(E) = -0.00180107E+0.82402941
	\label{eq12}
\end{equation}

Where the R-Score in this model is 0.856 indicating good model accuracy.

\begin{figure}[h!]
		\centering 
		\includegraphics[width=\linewidth]{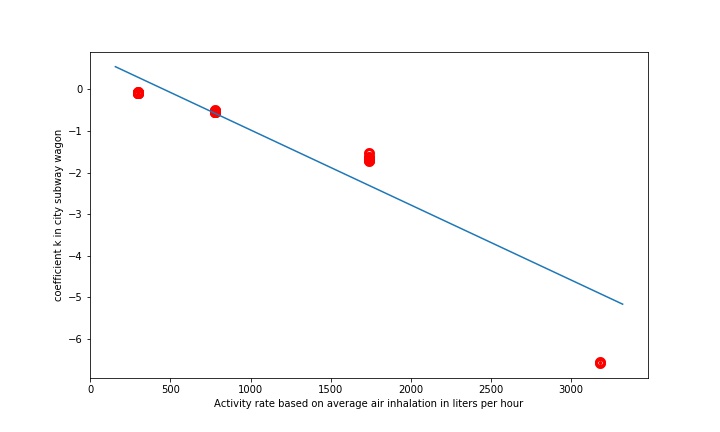}
		\caption{The value of $k(E)$ according to the amount of activity along with the estimated linear regression model in the subway wagon}
		\label{fig:fig1_4}
\end{figure}

Figure \eqref{fig:fig1_4} shows the relationship between k(E) versus activity level and the estimated linear regression model in subway wagons.

Finally, by substituting the k(E) value from equation \eqref{eq12} in equation \eqref{eq10}, we arrive at equation \eqref{eq13} which represents the probability of contracting disease in subway wagons.
\begin{equation}
	f(m,n,z,E)=1-e^{\sum_{j=1}^m{\sum_{i=1}^n{\frac{(-0.00180107E+0.82402941)z}{mr_{ij}^2}}}}
	\label{eq13}
\end{equation}

\subsubsection*{Calculating Coefficient k(E) and Probability of Contracting Disease in BRT}
The dimensions of Tehran's BRT buses are $17.9 \times 2.55$ meters \cite{BRTDimensions}. We consider this environment as a space with semi-open windows, average ventilation, and a 45-square-meter area. The information in tables \eqref{tab10}, \eqref{tab11}, and \eqref{tab12} has been calculated for three conditions of sitting, resting, mild or moderate activity in the BRT environment \cite{covid-19}. 

\begin{table} [!h]
  \centering
      \caption{Coefficient
    $k$
  in a room with a size of 45 square meters and the activity of sitting or resting for different activity times}
    \label{tab10}
    \begin{tabular}{|c|c|c|}
    \hline
         \textbf{Time in hours} & \textbf{Percentage of infection} & \textbf{$k$} \\
    \hline  
         1 & 3 & $-0.101871$ \\
    \hline 
         2 & 6 & $-0.103471$ \\
    \hline 
         3 & 8 & $-0.092957$ \\
    \hline 
         4 & 11 & $-0.097437$ \\
    \hline 
         5 & 14 & $-0.100886$ \\
    \hline 
         6 & 17 & $-0.103864$ \\
    \hline 
         7 & 20 & $-0.106615$ \\
    \hline 
         8 & 23 & $-0.109267$ \\
    \hline 
         9 & 25 & $-0.106906$ \\
    \hline
         10 & 28 & $-0.109868$ \\
    \hline  
    \end{tabular}
\end{table}

\begin{table} [!h]
  \centering
      \caption{Coefficient
    $k$
   in a room with a size of 45 square meters and low activity for different activity times}
    \label{tab11}
    \begin{tabular}{|c|c|c|}
    \hline
         \textbf{Time in hours} & \textbf{Percentage of infection} & \textbf{$k$} \\
    \hline  
         1 & 18 & $-0.663721$ \\
    \hline 
         2 & 34 & $-0.694848$ \\
    \hline 
         3 & 46 & $-0.686947$ \\
    \hline 
         4 & 57 & $-0.705666$ \\
    \hline 
         5 & 65 & $-0.702228$ \\
    \hline 
         6 & 72 & $-0.709574$ \\
    \hline 
         7 & 78 & $-0.723431$ \\
    \hline 
         8 & 82 & $-0.716895$ \\
    \hline 
         9 & 86 & $-0.730631$ \\
    \hline
         10 & 89 & $-0.738225$ \\
    \hline  
    \end{tabular}
\end{table}

\begin{table} [!h]
  \centering
      \caption{Coefficient
    $k$
    In a room with a size of 45 square meters and moderated activity for different activity times}
    \label{tab12}
    \begin{tabular}{|c|c|c|}
    \hline
         \textbf{Time in hours} & \textbf{Percentage of infection} & \textbf{$k$} \\
    \hline  
         1 & 48 & $-2.187063$ \\
    \hline 
         2 & 74 & $-2.252650$ \\
    \hline 
         3 & 87 & $-2.274513$ \\
    \hline 
         4 & 93 & $-2.223480$ \\
    \hline 
         5 & 97 & $-2.345543$ \\
    \hline 
         6 & 98 & $-2.180633$ \\
    \hline 
         7 & 99 & $-2.200291$ \\
    \hline 
    \end{tabular}
\end{table}

Similar to the open environment and subway wagon cases, by comparing the values of the above three tables with the average air intake for different activities given in Table \eqref{tab5}, we have calculated coefficient k(E) for the BRT environment in Python. The correlation coefficient between coefficient K and activity level in this case is -0.997468. Also, using the linear regression model in Python, coefficient k(E) was calculated as follows:

\begin{equation}
	k(E) = -0.00149112E+0.38875338
	\label{eq14}
\end{equation}

Where the r-value in this model is 0.995 indicating good model accuracy.

\begin{figure}[h!]
		\centering 
		\includegraphics[width=\linewidth]{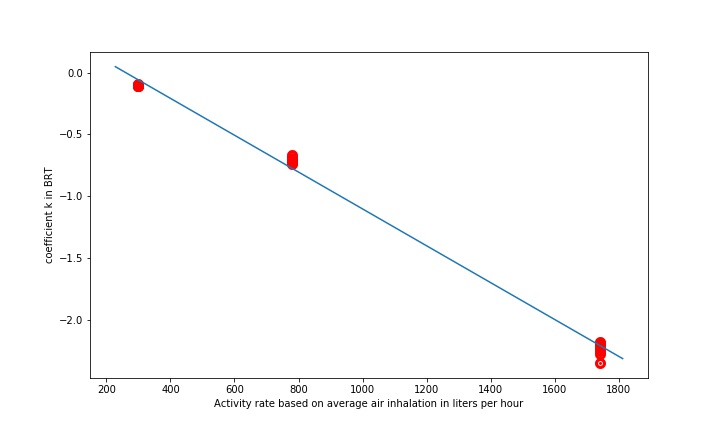}
		\caption{The value of $k(E)$ according to the amount of activity along with the estimated linear regression model in BRT}
		\label{fig:fig1_5}
\end{figure}

Figure \eqref{fig:fig1_5} shows the relationship between k(E) versus activity level and the estimated linear regression model in the BRT environment.

Finally, by substituting the k(E) value from equation \eqref{eq14} in equation \eqref{eq10}, we arrive at equation \eqref{eq15} which represents the probability of contracting disease in BRT.

\begin{equation}
	f(m,n,z,E)=1-e^{\sum_{j=1}^m{\sum_{i=1}^n{\frac{(-0.00149112E+0.38875338)z}{mr_{ij}^2}}}}
	\label{eq15}
\end{equation}

\subsection*{Calculating Coefficient k(E) and Probability of Contracting Disease in City Buses}
The dimensions of city buses in Tehran are $12 \times 2.55$ meters \cite{BusCapacity}. We consider this environment as a space with semi-open windows, average ventilation, and a 30-square-meter area. The information in tables \eqref{tab13}, \eqref{tab14}, and \eqref{tab15} has been calculated for three conditions of sitting, resting, mild or moderate activity in a city bus environment \cite{covid-19}.

\begin{table} [!h]
  \centering
      \caption{Coefficient
    $k$
   in a room with a size of 30 square meters and the activity of sitting or resting for different activity times}
    \label{tab13}
    \begin{tabular}{|c|c|c|}
    \hline
         \textbf{Time in hours} & \textbf{Percentage of infection} & \textbf{$k$} \\
    \hline  
         1 & 4 & $-0.136530$ \\
    \hline 
         2 & 8 & $-0.139435$ \\
    \hline 
         3 & 12 & $-0.142513$ \\
    \hline 
         4 & 16 & $-0.145782$ \\
    \hline 
         5 & 20 & $-0.149261$ \\
    \hline 
         6 & 24 & $-0.152976$ \\
    \hline 
         7 & 28 & $-0.156955$ \\
    \hline 
         8 & 32 & $-0.161231$ \\
    \hline 
         9 & 35 & $-0.160084$ \\
    \hline
         10 & 39 & $-0.165318$ \\
    \hline  
    \end{tabular}
\end{table}

\begin{table} [!h]
  \centering
      \caption{Coefficient
    $k$
  in a room with a size of 30 square meters and low activity for different activity times}
    \label{tab14}
    \begin{tabular}{|c|c|c|}
    \hline
         \textbf{Time in hours} & \textbf{Percentage of infection} & \textbf{$k$} \\
    \hline  
         1 & 26 & $-1.007049$ \\
    \hline 
         2 & 46 & $-1.030420$ \\
    \hline 
         3 & 61 & $-1.049740$ \\
    \hline 
         4 & 72 & $-1.064361$ \\
    \hline 
         5 & 80 & $-1.076556$ \\
    \hline 
         6 & 85 & $-1.057489$ \\
    \hline 
         7 & 90 & $-1.100145$ \\
    \hline 
         8 & 93 & $-1.111740$ \\
    \hline 
         9 & 95 & $-1.113251$ \\
    \hline
         10 & 96 & $-1.076556$ \\
    \hline  
    \end{tabular}
\end{table}

\begin{table} [!h]
  \centering
      \caption{Coefficient
    $k$
   in a room with a size of 30 square meters and moderated activity for different activity times}
    \label{tab15}
    \begin{tabular}{|c|c|c|}
    \hline
         \textbf{Time in hours} & \textbf{Percentage of infection} & \textbf{$k$} \\
    \hline  
         1 & 63 & $-3.325286$ \\
    \hline 
         2 & 87 & $-3.411769$ \\
    \hline 
         3 & 95 & $-3.339752$ \\
    \hline 
         4 & 98 & $-3.270949$ \\
    \hline 
    \end{tabular}
\end{table}

Similar to an open environment, subway wagons, and BRT, by comparing the values of the three tables above with the average air intake for different activities given in Table \eqref{tab5}, we have calculated the coefficient (E) for the city bus environment in Python. The correlation coefficient between coefficient K and activity level in this case is -0.997452. Also, using the linear regression model in Python, coefficient k(E) was calculated as follows:

\begin{equation}
	k(E) = -0.00220276E+0.56565911
	\label{eq18}
\end{equation}

Where the r-value in this model is 0.995 indicating good model accuracy.

\begin{figure}[h!]
		\centering 
		\includegraphics[width=\linewidth]{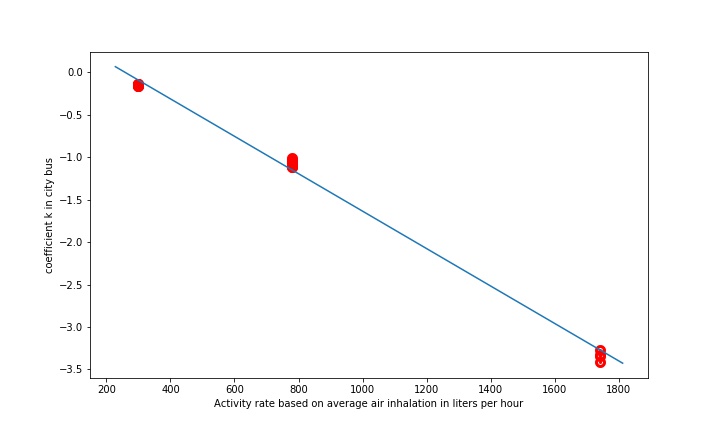}
		\caption{The value of $k(E)$ according to the amount of activity along with the estimated linear regression model in the city bus}
		\label{fig:fig1_6}
\end{figure}

Figure \eqref{fig:fig1_6} shows the relationship between k(E) versus activity level and the estimated linear regression model in the city bus environment.

Finally, by substituting the k(E) value from equation \eqref{eq18} in equation \eqref{eq10}, we arrive at equation \eqref{eq19} which represents the probability of contracting disease in the city bus environment.

\begin{equation}
	f(m,n,z,E)=1-e^{\sum_{j=1}^m{\sum_{i=1}^n{\frac{(-0.00220276E+0.56565911)z}{mr_{ij}^2}}}}
	\label{eq19}
\end{equation}

\subsection*{Calculating Coefficient K and Probability of Contracting Disease in Cars}
In a car without a proper ventilation system, it only takes 6.38 minutes for an infected driver to transmit the coronavirus to another passenger \cite{sarhan2022numerical}. According to the information in this paper, the location of the infected passenger and ventilation system affect the disease contraction time, such that if front-seat passengers are infected, the probability of disease transmission to other passengers is higher compared to when rear-seat passengers are infected. In this paper, disease contraction time is calculated for 4 different cases of the air ventilation system. Below, assuming the passenger car under study has a relatively good ventilation system with 4 meters per 4-meter-per-second ventilation rate, we calculate the average disease contraction time and subsequently the coefficient.

With the above assumption, if the driver or front seat passenger is infected, on average it takes about 20 minutes for other passengers to become infected. If the rear seat passenger behind the driver is infected, on average it takes about 17 minutes for other passengers to become infected. Finally, if the other passenger is infected, on average it takes about 37 minutes for others to become infected. With this calculation, the average infection time by taking the average of the above times is 23.5 minutes.

In this case, to calculate coefficient K in equation \eqref{eq10}, we set the time parameter (z) equal to 23.5/60 and consider the probability of contracting the disease equal to 0.99. Thus, we arrive at equation \eqref{eq16}:

\begin{equation}
	1-e^{\sum_{j=1}^m{\sum_{i=1}^n{\frac{6.38k}{60mr_{ij}^2}}}}=0.99
	\label{eq16}
\end{equation}

Assuming there is one infected person in the car and the average distance between two people is 4 meters (this distance was calculated in Chapter 3 and is given in Table 2), we arrive at the following relationship:

\begin{equation}
	1-e^{\frac{23.5k}{60\times(0.48^2)}}=0.99
	\label{eq17}
\end{equation}

By solving equation \eqref{eq17}, the K value in this case is equal to -2.72948. Finally, by substituting the K value in equation \eqref{eq20}, the probability of contracting disease in passenger cars is obtained as follows:

\begin{equation}
	f(m,n,z)=1-e^{\sum_{j=1}^m{\sum_{i=1}^n{\frac{-2.729480z}{mr_{ij}^2}}}}
	\label{eq20}
\end{equation}

\section*{Choose the best route}

In this chapter, the model defined in the previous chapter is applied to various paths between a specified origin and destination. We determine the optimal path considering the minimum probability of contracting the disease. To achieve this, routing algorithms are employed using the navigation applications "Neshan" and "Balad" for routes between Sadeghiyeh Square and Amir Kabir University in Tehran city, specifically at 18:00 on a weekday. The general routing algorithm is presented. Then, the number of infected individuals and the average distance between individuals on different paths are calculated in the general case. Next, a general relationship for the probability of contracting the disease in each possible environment is redefined based on the equations derived from the previous chapter. Finally, the specific relationship for disease contraction in different environments is determined. The probability of contracting the disease using the recommended routes in the "Neshan" and "Balad" applications is computed.
\subsection*{Routing Algorithm}
According to the equations $\eqref{eq11}$, $\eqref{eq13}$, $\eqref{eq15}$, $\eqref{eq19}$ and $\eqref{eq20}$ from the previous chapter, the probability of contracting the disease in different environments can be calculated. It is assumed that the distance from the individual under consideration to infected individuals remains constant throughout the entire route. With this assumption, the relationship \eqref{eq10} is expressed as:
$f(x,y,z)=1-e^{{\sum_{i=1}^n{\frac{kz}{r_{i}^2}}}}$
As a result, the derived equations for different routes are independent of the parameter $m$, which represents the length of the environment. Additionally, assuming that the distance between infected individuals and the individual under consideration is equal to the average distance at any given moment, these relationships can be summarized for pedestrian, subway, BRT, and city bus environments according to the equation  $\eqref{eq2_3}$.

\begin{equation}
	1-e^{\frac{(k_1E+k_2)nz}{r_{mean}^2}}
	\label{eq2_3}
\end{equation}	
Table $\eqref{tab2_4}$ provides $k_1$ and $k_2$ coefficients.
\begin{table} [!h]
  \centering
      \caption{$k_1$ and $k_2$ coefficients to calculate the probability of getting a disease on walking, subway, BRT and city buses}
    \label{tab2_4}
    \begin{tabular}{|c|c|c|}
    \hline
         \textbf{environment} & \textbf{$k_1$} & \textbf{$k_2$} \\
    \hline  
         walking & $-0.00143853$ & $0.71455401$  \\
    \hline 
         subway & $-0.00180107$ & $0.82402941$  \\
    \hline 
         BRT & $-0.00149112$ & $0.38875338$  \\
    \hline 
         city bus & $-0.00220276$ & $0.56565911$  \\
    \hline
    \end{tabular}
\end{table}

The same relationship applies to the passenger car, formulated as $\eqref{eq2_4}$.

\begin{equation}
	1-e^{\frac{-2.729480nz}{r_{mean}^2}}
	\label{eq2_4}
\end{equation}

The input parameters for these functions are as follows:
\begin{itemize}
	\item
	z: The time the individual under consideration spends in the environment.
	\item
	E: The activity level is equivalent to the air inhalation rate in liters per hour. Table $\eqref{tab5}$ can be used without an exact value.
	\item
	n: The number of the infected. To initialize this parameter, the probability that an individual in the community is infected must first be obtained. For example, based on calculations provided in the following section, this value was calculated to be 0.8656
	\item
	$r_{mean}$: Assuming the environment is rectangular, the average distance between two individuals in the environment can be calculated according to the equation \eqref{eq2_2}.
\end{itemize}
	
The following are the steps of the algorithm for finding the optimal route. This is assuming that the routing application has suggested several routes. Each route comprises walking, subway, BRT, city bus, and car segments. The objective is to find the route with the lowest probability of contracting the disease.
Routing Algorithm:
\begin{enumerate}
	\item
    For each route segment, calculate the values of parameters n (according to the hour of the day under consideration), $r_{mean}$ (according to the length and width of the environment), E, and z.
    \item
    Calculate the probability of contracting the disease for each route segment based on the equations $\eqref{eq2_3}$ and $\eqref{eq2_4}$.
    \item
    After substituting the probabilities obtained in the previous step into the equation $\eqref{eq0}$, we calculate the probability of contracting the disease for the entire route.
    \item
    We repeat the above steps for the other suggested routes.
    \item
    The route with the lowest probability of contracting the disease is recommended as the selected route.
\end{enumerate}

\subsection*{Number of Infected Individuals and Average Distance}
To calculate the probability of contracting the disease on each route, it is necessary to know the average number of infected individuals and the average distance between individuals in each environment under consideration. During the peak of the COVID-19 period from September 1, 2021, to September 21, 2021, the total number of COVID-19 patients reported in Iran was 5,295,786, of which 4,568,236 individuals recovered.\cite{covidIranShahrivar1400} It can be inferred that there are 727,550 infected individuals in society on this date. Iran's population in 1400 is estimated at 84,055,000 individuals.\cite{IranPopulation} Based on this calculation, the probability of COVID-19 in Iranian society during the pandemic's peak is 0.8656
\begin{equation}
	Average\;Distance = \frac{1}{15}(\frac{L_w^3}{L_h^2}+\frac{L_h^3}{L_w^2}+d(3-\frac{L_w^2}{L_h^2}-\frac{L_h^2}{L_w^2})+\frac{5}{2}(\frac{L_h^2}{L_w}log{\frac{L_w+d}{L_h}}+\frac{L_w^2}{L_h}log{\frac{L_h+d}{L_w}})) 
	\label{eq2_2}	
\end{equation}
$,where\; d=\sqrt{L_w^2+L_h^2}$

\begin{figure}[h!]
		\centering 
		\includegraphics[width=\linewidth]{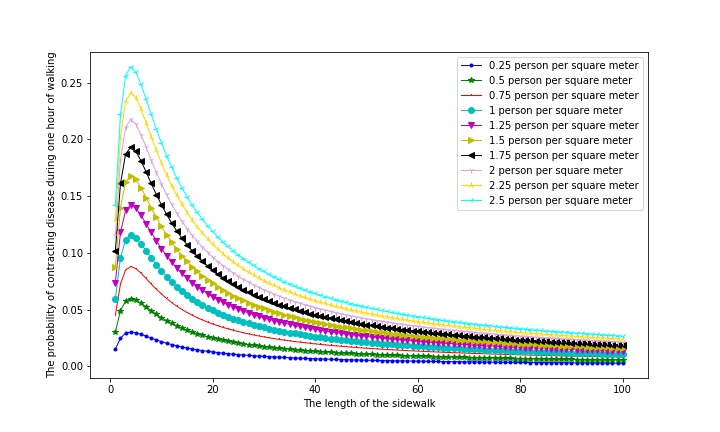}
		\caption{The probability of getting a disease in one hour of walking in an environment with a width of 4 meters according to the length of the environment and for different densities of the environment}
		\label{fig:fig2_0}
\end{figure}

For subway, city buses, BRT, and car routes, the dimensions of the environment are specified; therefore, according to equation \eqref{eq2_2}, the average distance can be easily calculated. However, since the environment is open to pedestrian routes, the environment dimensions are not specified. Therefore, to examine the suitable average distance in pedestrian mode, a graph $\eqref{fig:fig2_0}$ is plotted. In this graph, the probability of contracting the disease, according to equation \eqref{eq2_3}, is shown based on the length of the pedestrian environment and for different densities of individuals per square meter. This graph assumes that a person spends one hour in an environment with a width of 4 meters. 

According to this graph, the highest probability of contracting the disease exists in a square environment with a pedestrian length greater than 4 meters. However, in environments with low congestion, from approximately 20 meters onwards, the probability of contracting the disease is almost constant. Therefore, with this assumption, we have considered the length of the pedestrian environment as 20 meters. With these specifications and by substituting the length and width of different environments into equation \eqref{eq2_2}, the average distance in different environments is obtained, as shown in Table \eqref{tab2_1}.

\begin{table} [!h]
  \centering
      \caption{The average distance between two people in meters in different environments}
    \label{tab2_1}
    \begin{tabular}{|c|c|c|c|}
    \hline
         \textbf{environment} & \textbf{length of environment} & \textbf{width of environment} & \textbf{average distance} \\
    \hline  
        walking & 20 & 12 & $4.95$ \\
    \hline 
         subway & $19.52$ & $2.6$ & $4.75$ \\
    \hline 
         BRT & $17.9$ & $2.55$ & $4.36$ \\
    \hline 
         city bus & $2.55$ & $12$ & $2.98$ \\
    \hline  
         car & $1.5$ & $1.2$ & $0.48$ \\
    \hline 
    \end{tabular}
\end{table}

The total number of individuals in each environment is determined based on the capacity of each mode of transportation \cite{MetroCapacity} \cite{BRTDimensions} because, during peak traffic hours, vehicles operate at their maximum capacity, as shown in Table \eqref{tab2_2}. If there is no peak traffic during rush hours, the number of individuals is lower. With these considerations, the final relationship depends on routing time during the day. As mentioned earlier, the number of infected individuals is obtained by multiplying the total number of individuals by 0.008656.

\begin{table} [!h]
  \centering
      \caption{The total number of people and infected people in different environments}
    \label{tab2_2}
    \begin{tabular}{|c|c|c|}
    \hline
         \textbf{environment} & \textbf{total number of people} & \textbf{number of infected people} \\
    \hline  
        walking & 40 & $0.34624$  \\
    \hline 
         subway & 180 & $1.55808$ \\
    \hline 
         BRT & 150 & $1.2984$ \\
    \hline 
         city bus & 80 & $0.69248$ \\
    \hline  
         car & 4 & $0.034624$ \\
    \hline 
    \end{tabular}
\end{table}

\subsection*{Probability Equations for Different Routes}
In this section, the probability function of contracting the disease is calculated for the specific case of the route between Sadeghiyeh Square and Amir Kabir University at 18:00 on a weekday. The calculations are performed during peak traffic hours and are based on the following assumptions:
\begin{itemize}
	\item
	The average time between each subway and city bus station is 3 minutes.
	\item
	The activity level for pedestrian routes is assumed to be average, and E is 1740.
	\item
	Subway, BRT, and city bus routes are assumed to be low, and E is 780.
    \item
    The average walking speed is assumed to be 5 kilometers per hour.
\end{itemize}

Table $\eqref{tab2_3}$ summarizes routing algorithm equations for different environments.
\begin{table} [!h]
  \centering
      \caption{Probability equations for different environments}
    \label{tab2_3}
    \begin{tabular}{|c|c|c|c|c|c|}
    \hline
         \textbf{environment} & \textbf{$n$} & \textbf{$E$}
         & \textbf{$r_{mean}$} & \textbf{$f(n,E,r_{mean},z)$} & \textbf{$f(z)$} \\
    \hline  
        walking & $0.34656$ & 1740 & $4.95$ & 
         $1-e^{\frac{(-0.00143853E+0.71455401)nz}{r_{mean}^2}}$ &
         $1-e^{-0.025299z}$  \\
	\hline  
         subway & $1.55808$ & 780 & $4.75$ & 
         $1-e^{\frac{(-0.00180107E+0.82402941)nz}{r_{mean}^2}}$ &
         $1-e^{-0.040155z}$  \\ 
         \hline  
         BRT & $1.2984$ & 780 & $4.36$ & 
         $1-e^{\frac{(-0.00149112E+0.38875338)nz}{r_{mean}^2}}$ &
         $1-e^{-0.052831z}$  \\ 
         \hline  
         city bus & $0.69248$ & 780 & $2.98$ & 
         $1-e^{\frac{(-0.00220276E+0.56565911)nz}{r_{mean}^2}}$ &
         $1-e^{-0.089912z}$  \\ 
         \hline  
         car & $0.034624$ & - & $0.48$ & 
         $1-e^{\frac{-2.729480nz}{r_{mean}^2}}$ &
         $1-e^{-0.407105z}$  \\ 
         \hline  
    \end{tabular}
\end{table}

\subsection*{Routing in the "Neshan" Application}
This section specifies routes between Sadeghiyeh Square and Amir Kabir University in Tehran at 18:00 on a weekday in the "Neshan" application. Six different routes are suggested, and the probability of contracting the disease is calculated for each route.
\subsubsection*{First Pedestrian Route}
\begin{figure}[h!]
		\centering 
		\includegraphics[width=\linewidth]{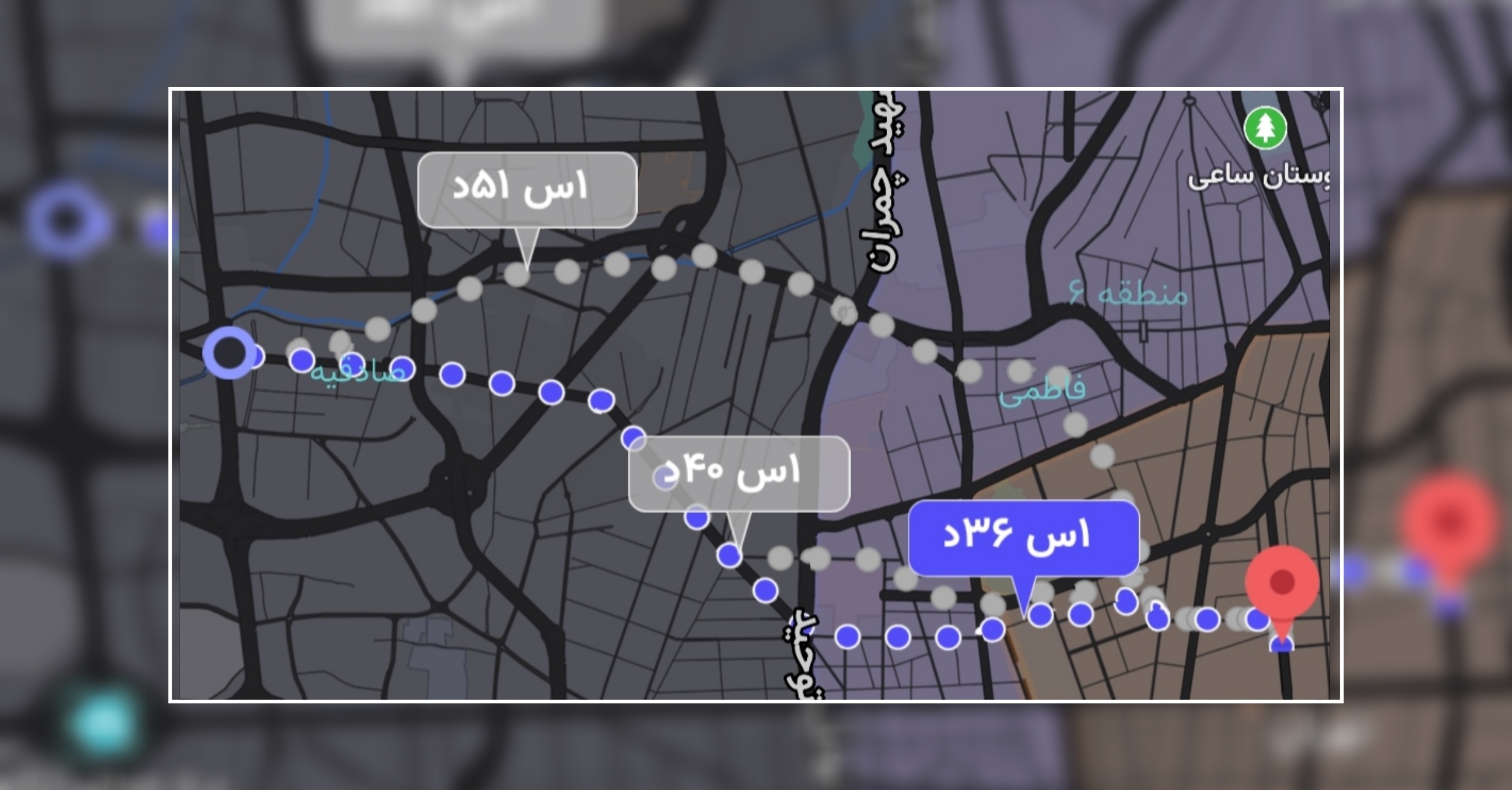}
		\caption{First route in "Neshan" application between Sadeghiyeh Square and Amir Kabir University}
		\label{fig:fig2_1}
\end{figure}

Figure \eqref{fig:fig2_1} shows that this route is 8.2 kilometers long and takes 1 hour and 36 minutes. Since the entire route is on foot and equivalent to an open environment, to calculate the probability of contracting the disease, we substitute the time into the function related to the pedestrian route in Table \eqref{tab2_3}.

\begin{equation}
	f_1=1-e^{-0.025299\times1.6}=0.0397
	\label{eq2_1}
\end{equation}

 As a result, the probability of contracting the disease on the first route is 3.97

\subsubsection*{Second Pedestrian and City Bus Route}
\begin{figure}[h!]
		\centering 
		\includegraphics[width=\linewidth]{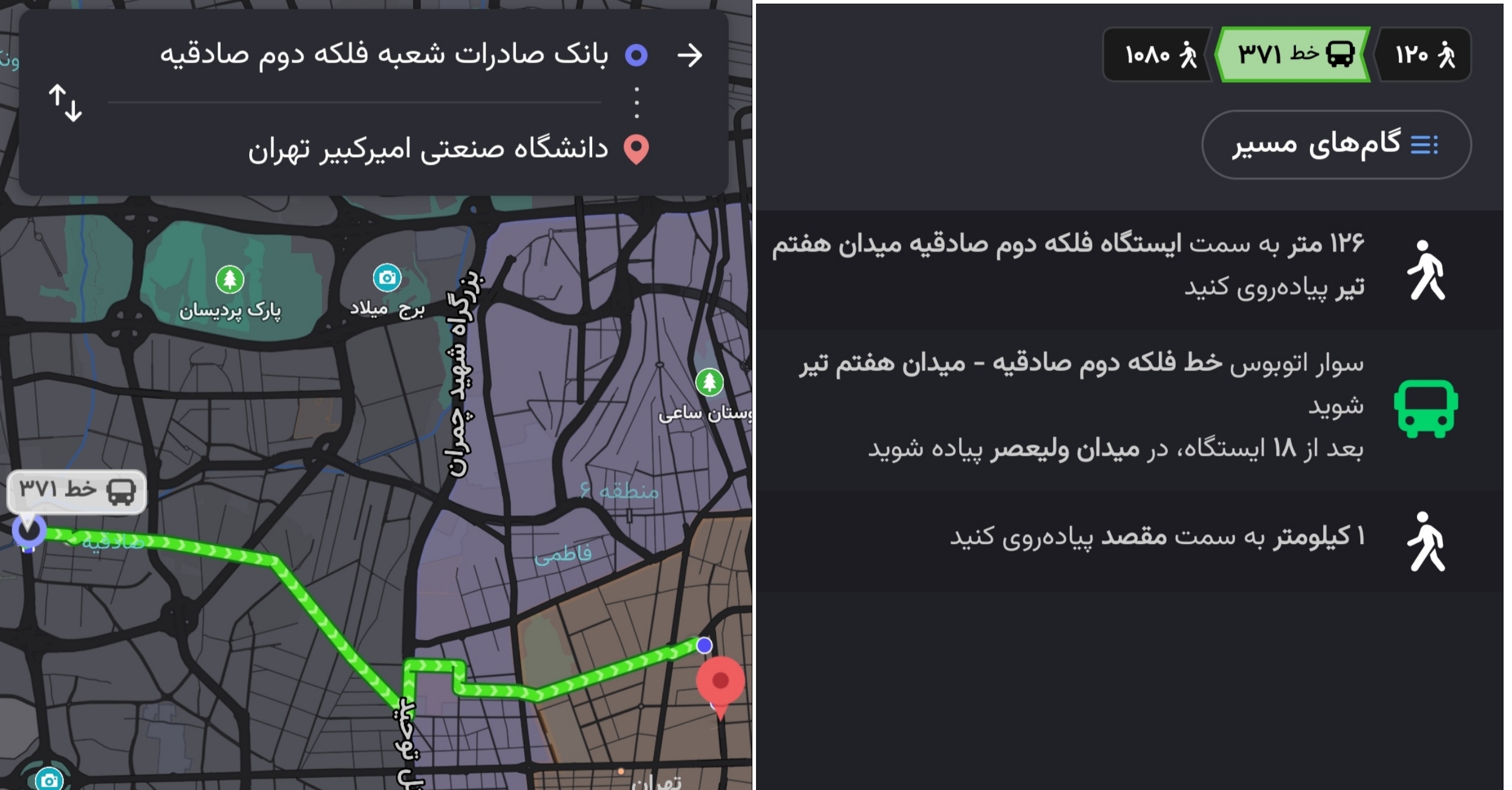}
		\caption{Second route in "Neshan" application between Sadeghiyeh Square and Amir Kabir University}
		\label{fig:fig2_2}
\end{figure}

Figure \eqref{fig:fig2_2} shows that the individual under consideration walks 126 meters, boards 18 city bus stops and walks 1080 meters. First, the probability of contracting the disease in each route segment is calculated.

\begin{equation}
	f_{21}=1-e^{-0.025299\times\frac{126}{5000}}=0.0006
\end{equation}
\begin{equation}
	f_{22}=1-e^{-0.089912\times\frac{18\times3}{60}}=0.0777
\end{equation}
\begin{equation}
	f_{23}=1-e^{-0.025299\times\frac{1080}{5000}}=0.0054
\end{equation}

Using equation \eqref{eq0}, the final probability of contracting the disease for the entire route is calculated as follows.

\begin{equation}
	f_2=1-(1-f_{21})(1-f_{22})(1-f_{23})=0.0833
\end{equation}

As a result, the probability of contracting the disease on the second route is 8.33

\subsubsection*{Third Pedestrian and City Bus Route}
\begin{figure}[h!]
		\centering 
		\includegraphics[width=\linewidth]{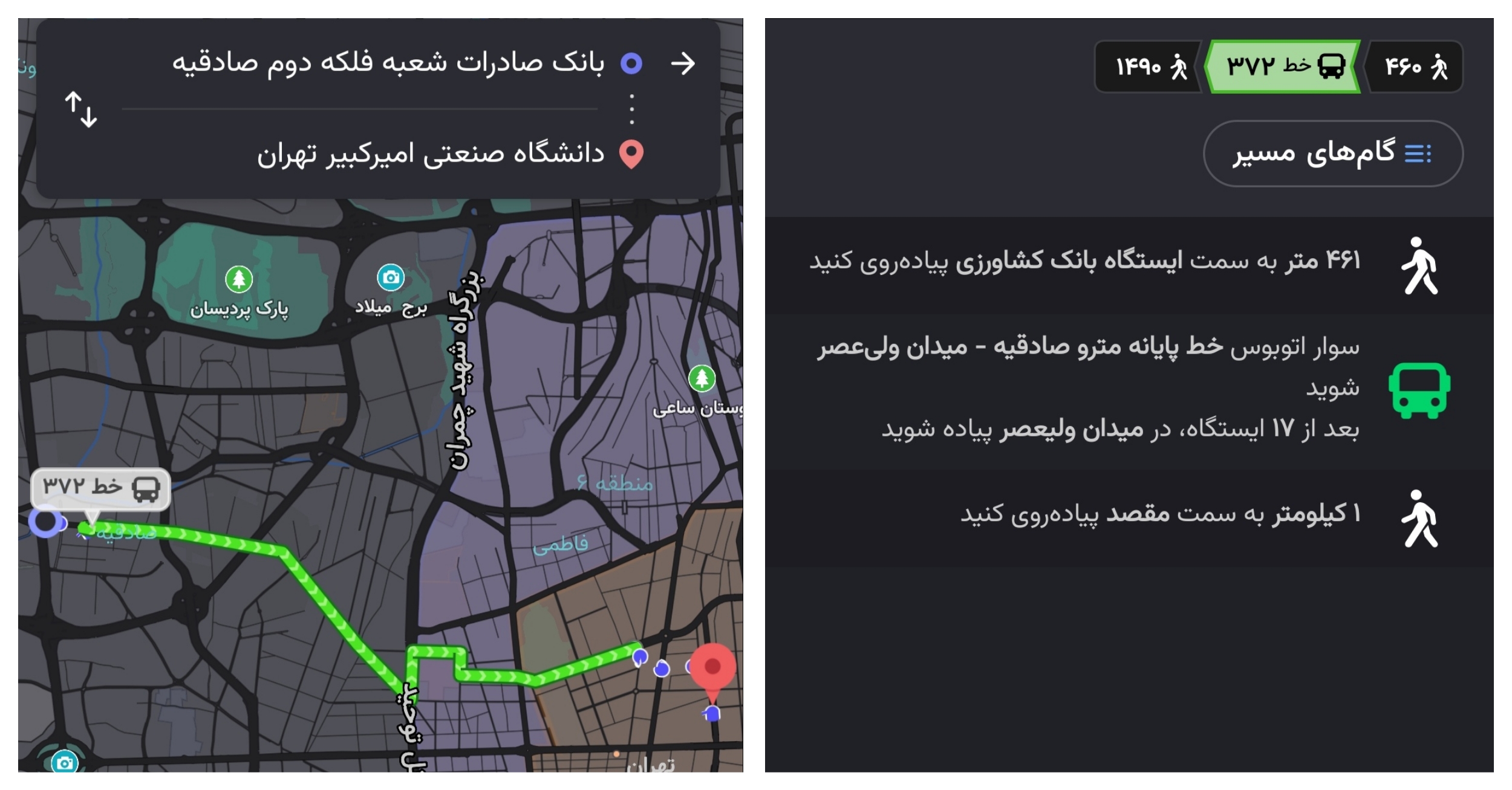}
		\caption{Third route in "Neshan" application between Sadeghiyeh Square and Amir Kabir University}
		\label{fig:fig2_3}
\end{figure}

Figure \eqref{fig:fig2_3} shows the individual under consideration walks 461 meters, boards 17 city bus stops, and finally walks 1490 meters. First, the probability of contracting the disease on each route segment is calculated.

\begin{equation}
	f_{31}=1-e^{-0.025299\times\frac{461}{5000}}=0.0023
\end{equation}
\begin{equation}
	f_{32}=1-e^{-0.089912\times\frac{17\times3}{60}}=0.0736
\end{equation}
\begin{equation}
	f_{33}=1-e^{-0.025299\times\frac{1490}{5000}}=0.0075
\end{equation}

Using equation \eqref{eq0}, the final probability of contracting the disease for the entire route is calculated as follows.

\begin{equation}
	f_3=1-(1-f_{31})(1-f_{32})(1-f_{33})=0.0827
\end{equation}

As a result, the probability of contracting the disease on the third route is 8.27

\subsubsection*{Fourth Pedestrian, BRT and Subway Route}
\begin{figure}[h!]
		\centering 
		\includegraphics[width=\linewidth]{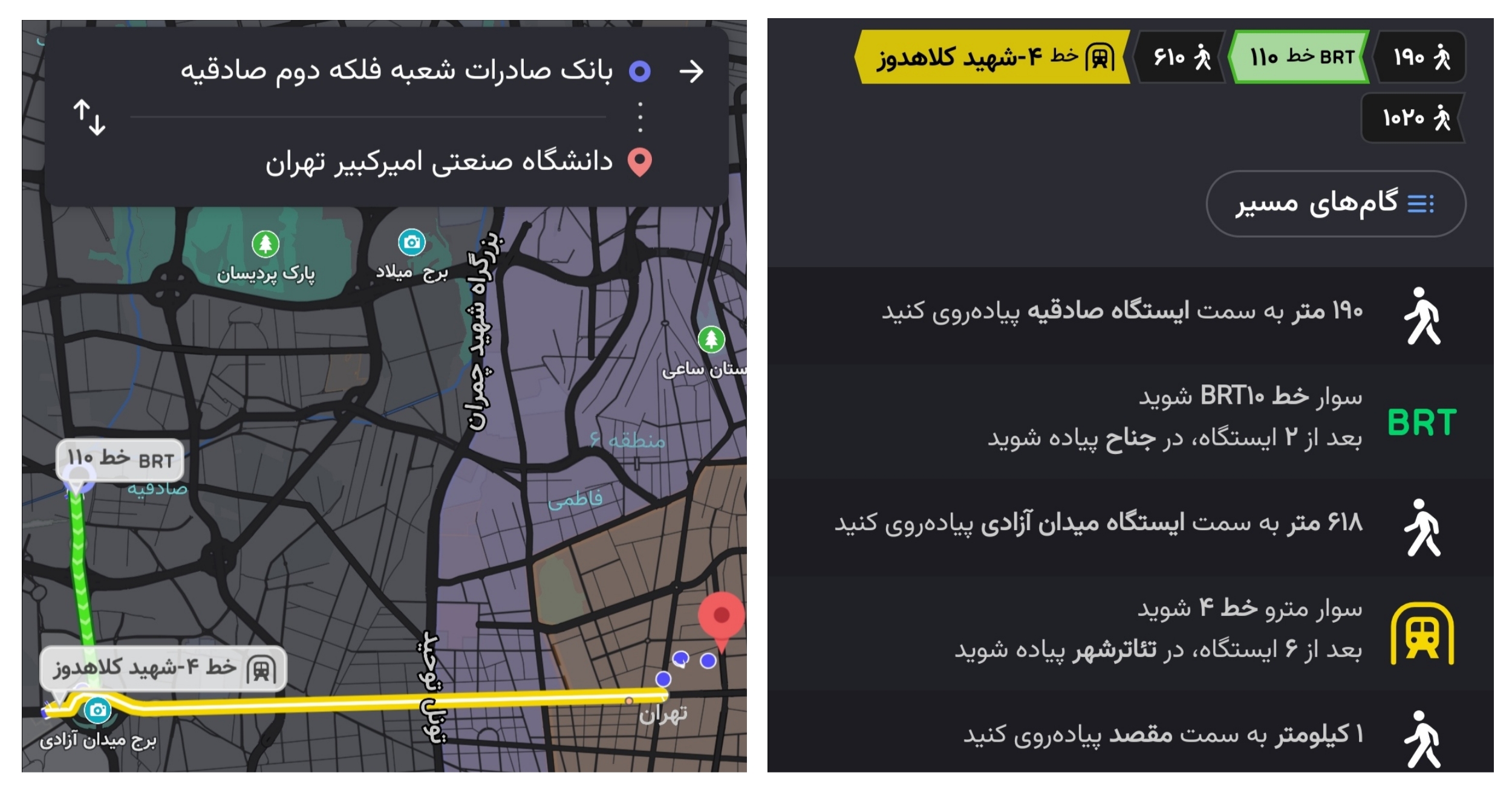}
		\caption{Fourth route in "Neshan" application between Sadeghiyeh Square and Amir Kabir University}
		\label{fig:fig2_4}
\end{figure}

According to Figure \eqref{fig:fig2_4}, the individual under consideration initially walks 190 meters, then covers 2 BRT stops, 618 meters, boards six subway stops, and finally walks 1020 meters. First, the probability of contracting the disease in each route segment is calculated.

\begin{equation}
	f_{41}=1-e^{-0.025299\times\frac{190}{5000}}=0.0010
\end{equation}
\begin{equation}
	f_{42}=1-e^{-0.052831\times\frac{2\times3}{60}}=0.0053
\end{equation}
\begin{equation}
	f_{43}=1-e^{-0.025299\times\frac{618}{5000}}=0.0031
\end{equation}
\begin{equation}
	f_{44}=1-e^{-0.040155\times\frac{6\times3}{60}}=0.0120
\end{equation}
\begin{equation}
	f_{45}=1-e^{-0.025299\times\frac{1020}{5000}}=0.0051
\end{equation}

Using the equation \eqref{eq0}, the final probability of contracting the disease for the entire route is calculated as follows.

\begin{equation}
	f_4=1-(1-f_{41})(1-f_{42})(1-f_{43})(1-f_{44})(1-f_{45})=0.0262
\end{equation}

As a result, the probability of contracting the disease on the fourth route is 2.62

\subsubsection*{Fifth Pedestrian, City Bus and BRT Route}

\begin{figure}[h!]
		\centering 
		\includegraphics[width=\linewidth]{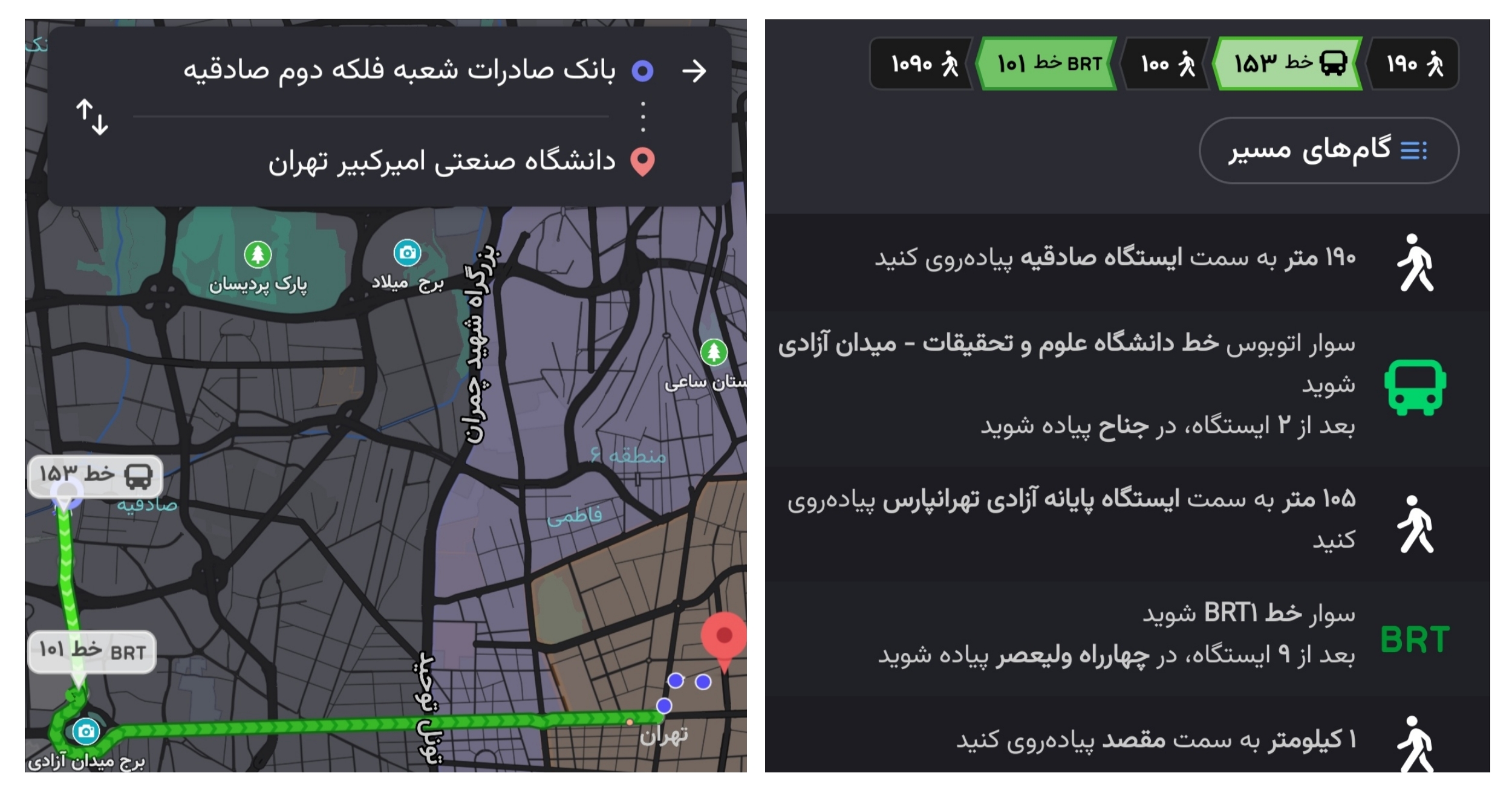}
		\caption{Fifth route in "Neshan" application between Sadeghiyeh Square and Amir Kabir University}
		\label{fig:fig2_5}
\end{figure}

According to Figure \eqref{fig:fig2_5}, the individual under consideration initially walks 190 meters, then covers two city bus stops, walks 105 meters, boards 9 BRT stops, and finally walks 1090 meters. First, the probability of contracting the disease in each route segment is calculated.

\begin{equation}
	f_{51}=1-e^{-0.025299\times\frac{190}{5000}}=0.0010
\end{equation}
\begin{equation}
	f_{52}=1-e^{-0.089912\times\frac{2\times3}{60}}=0.0090
\end{equation}
\begin{equation}
	f_{53}=1-e^{-0.025299\times\frac{105}{5000}}=0.0005
\end{equation}
\begin{equation}
	f_{54}=1-e^{-0.052831\times\frac{9\times3}{60}}=0.0235
\end{equation}
\begin{equation}
	f_{55}=1-e^{-0.025299\times\frac{1090}{5000}}=0.0055
\end{equation}

Using the equation \eqref{eq0}, the final probability of contracting the disease for the entire route is calculated as follows.

\begin{equation}
	f_5=1-(1-f_{51})(1-f_{52})(1-f_{53})(1-f_{54})(1-f_{55})=0.0390
\end{equation}

As a result, the probability of contracting the disease on the fifth route is 3.90

\subsubsection*{Sixth Car Route}
\begin{figure}[h!]
		\centering 
		\includegraphics[width=\linewidth]{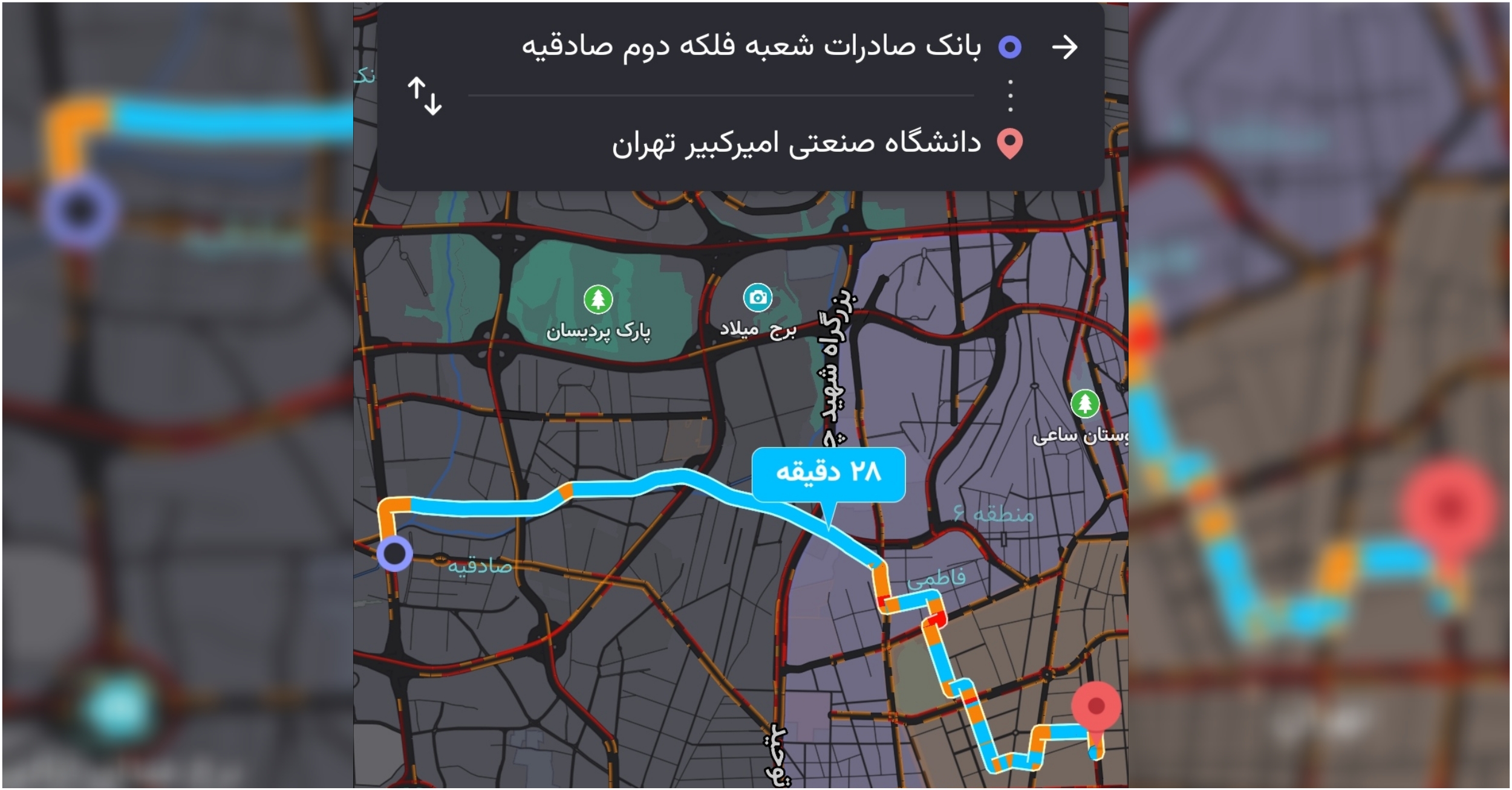}
		\caption{Sixth route in "Neshan" application between Sadeghiyeh Square and Amir Kabir University}
		\label{fig:fig2_6}
\end{figure}

Figure \eqref{fig:fig2_6} shows that the individual under consideration has been in the car for 28 minutes. This calculation calculates the probability of contracting the disease as follows.

\begin{equation}
	f_6=1-e^{-0.407105\times\frac{28}{60}}=0.1730
\end{equation}

As a result, the probability of contracting the disease on the sixth route is 17.30

In conclusion, among the recommended routes in the "Neshan" application, the fourth route, a combination of pedestrian, subway, and BRT routes, is the best route for the probability of contracting the disease, having the lowest likelihood.

\subsection*{Routing in "Balad" Application}
In this section, we have calculated the probability of contracting the disease for five different routes suggested in the "Balad" application. These routes are suggested for weekdays at 18:00.

\subsubsection*{First Pedestrian Route}
Figure $\eqref{tab2_7}$  recommends a pedestrian route of 8.4 kilometers, and 1 hour and 45 minutes is recommended. Since the entire route is pedestrian and equivalent to an open environment, we use time substitution in the function related to the pedestrian route in Table $\eqref{tab2_3}$  to calculate the probability of contracting the disease. 
\begin{figure}[h!]
		\centering 
		\includegraphics[width=\linewidth]{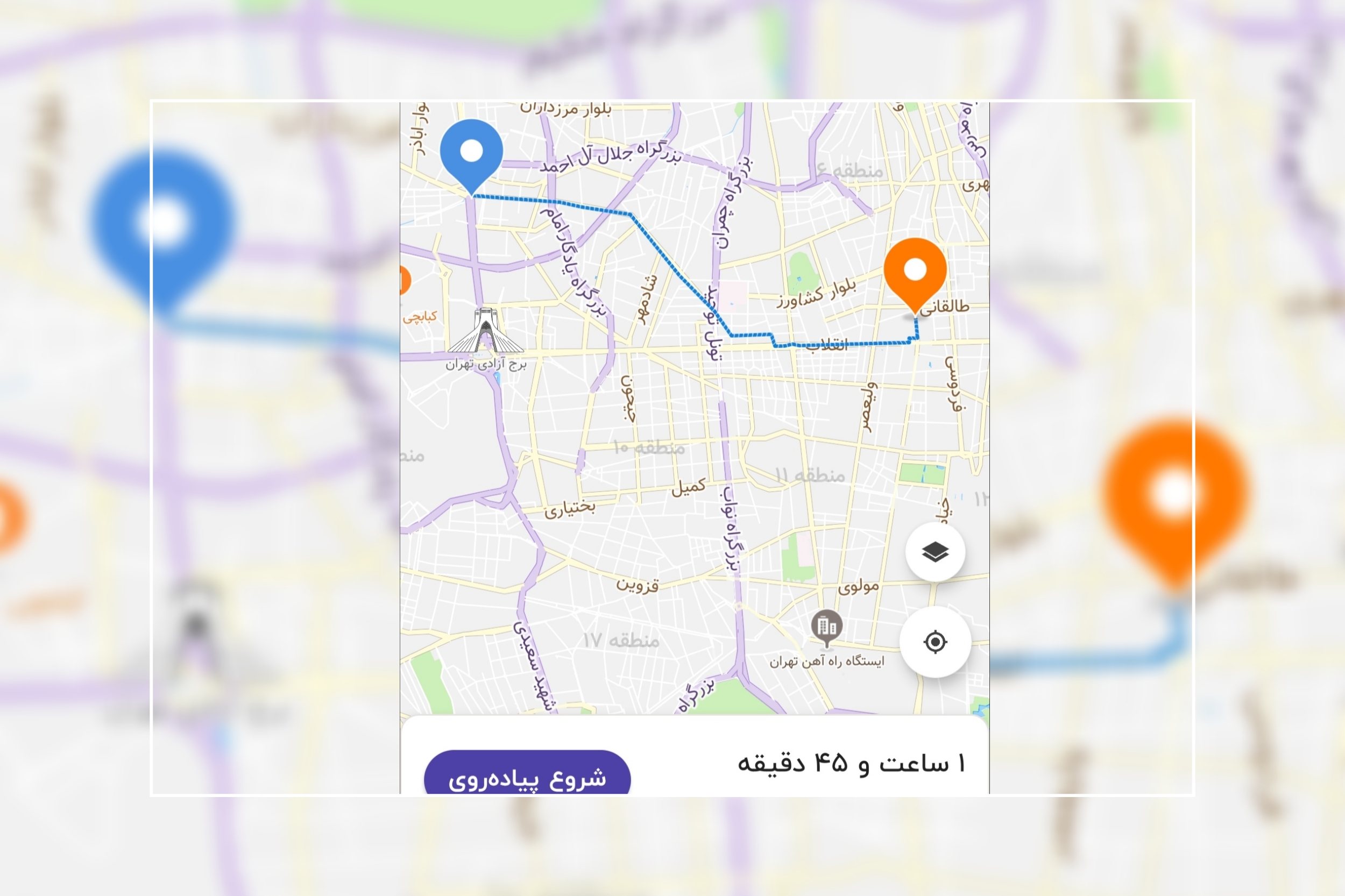}
		\caption{First route in "Balad" application between Sadeghiyeh Square and Amir Kabir University}
		\label{fig:fig2_7}
\end{figure} 

\begin{equation}
	f_1=1-e^{-0.025299\times1.6}=0.0433
\end{equation}

As a result, the probability of contracting the disease on the first route is 4.33

\subsubsection*{Second Pedestrian and City Bus Route}
\begin{figure}[h!]
		\centering 
		\includegraphics[width=\linewidth]{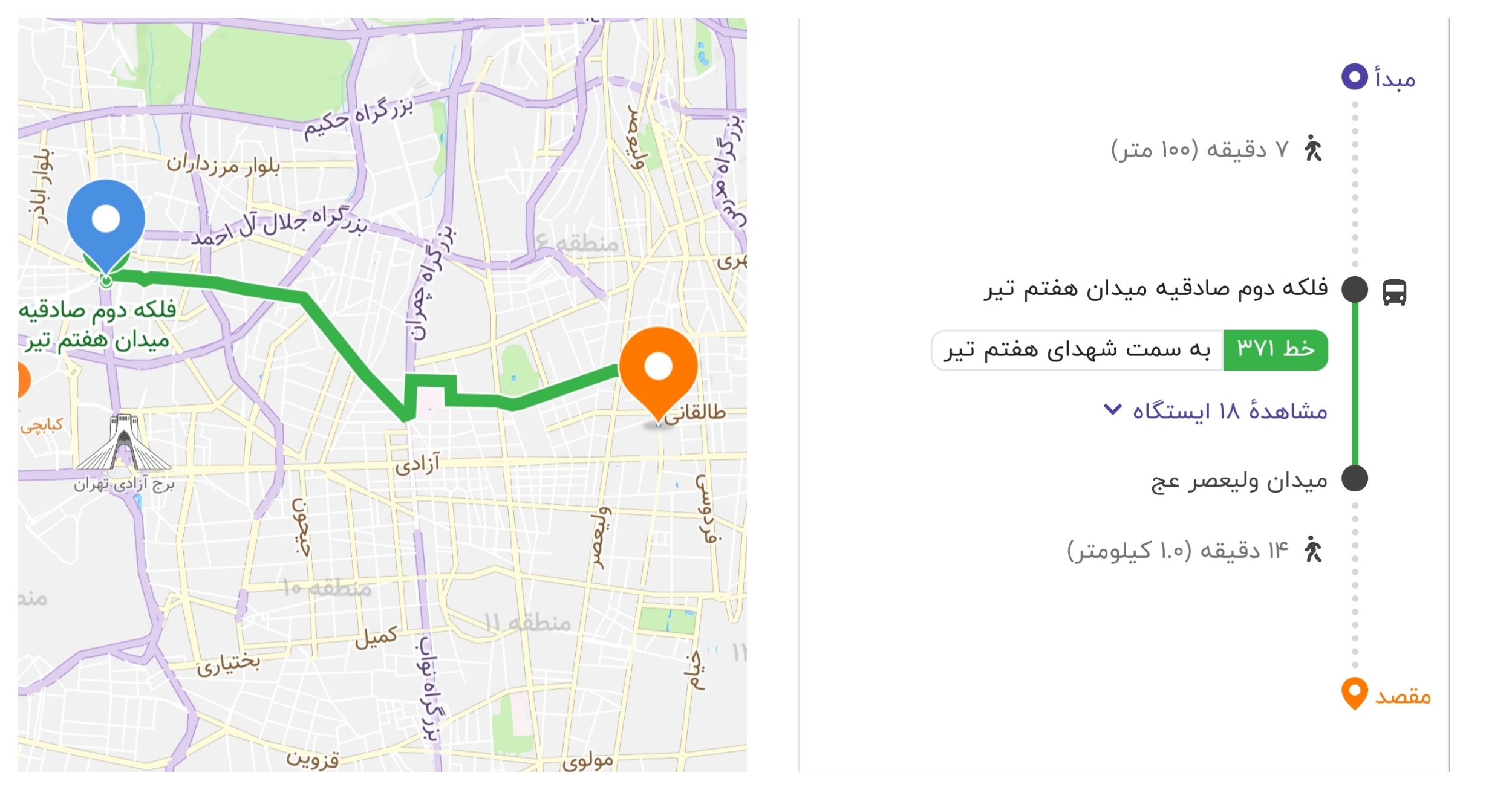}
		\caption{Second route in "Balad" application between Sadeghiyeh Square and Amir Kabir University}
		\label{fig:fig2_8}
\end{figure}

Figure \eqref{fig:fig2_8} shows the individual under consideration walks 100 meters, then boards 18 city bus stops, and walks another 1000 meters. Initially, the probability of contracting the disease in each part of the route was calculated. 

\begin{equation}
	f_{21}=1-e^{-0.025299\times\frac{100}{5000}}=0.0005
\end{equation}
\begin{equation}
	f_{22}=1-e^{-0.089912\times\frac{18\times3}{60}}=0.0777
\end{equation}
\begin{equation}
	f_{23}=1-e^{-0.025299\times\frac{1000}{5000}}=0.0050
\end{equation}

Finally, using the equation \eqref{eq0}, the ultimate probability of contracting the disease for the entire route is determined. 

\begin{equation}
	f_2=1-(1-f_{21})(1-f_{22})(1-f_{23})=0.0829
\end{equation}

As a result, the probability of contracting the disease on the second route is 8.29

\subsubsection*{Third Pedestrian and Car Route}
\begin{figure}[h!]
		\centering 
		\includegraphics[width=\linewidth]{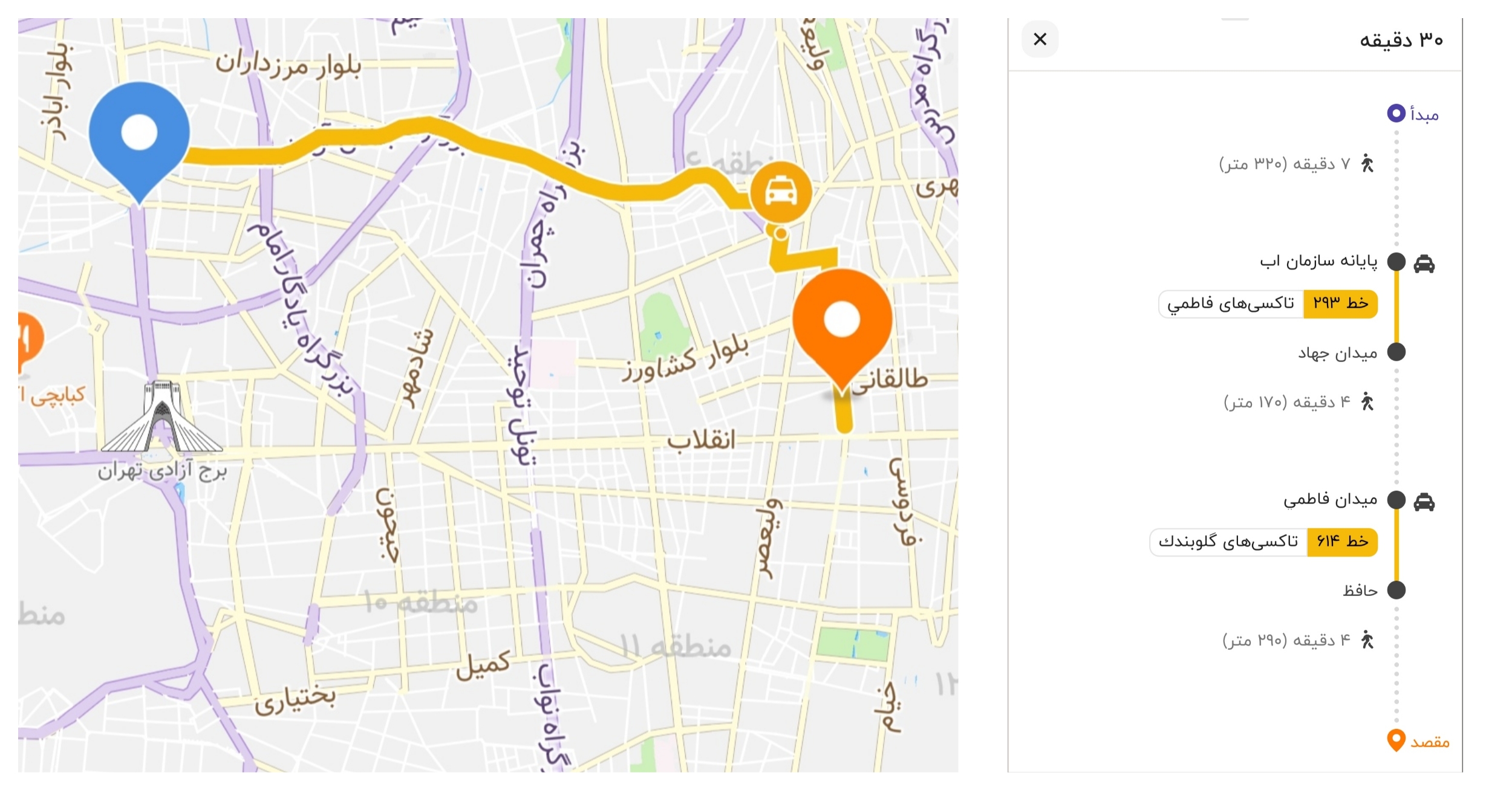}
		\caption{Third route in "Balad" application between Sadeghiyeh Square and Amir Kabir University}
		\label{fig:fig2_9}
\end{figure}

Figure \eqref{fig:fig2_9} shows that the individual under consideration walked for 7 minutes. Then, they were in a car for 8 minutes, followed by another 4 minutes of walking. Subsequently, they spent 7 minutes in the car and walked for 4 minutes. Initially, the probability of contracting the disease in each part of the route was calculated. 

\begin{equation}
	f_{31}=1-e^{-0.025299\times\frac{7}{60}}=0.0029
\end{equation}
\begin{equation}
	f_{32}=1-e^{-0.407105\times\frac{8}{60}}=0.0528
\end{equation}
\begin{equation}
	f_{33}=1-e^{-0.025299\times\frac{4}{60}}=0.0017
\end{equation}
\begin{equation}
	f_{32}=1-e^{-0.407105\times\frac{7}{60}}=0.0464
\end{equation}
\begin{equation}
	f_{33}=1-e^{-0.025299\times\frac{4}{60}}=0.0017
\end{equation}

Finally, using the equation \eqref{eq0}, the ultimate probability of contracting the disease for the entire route is determined. 

\begin{equation}
	f_3=1-(1-f_{31})(1-f_{32})(1-f_{33})(1-f_{43})(1-f_{53})=0.1025
\end{equation}

As a result, the probability of contracting the disease on the third route is 10.25

\subsubsection*{Fourth Pedestrian and Subway Route}
\begin{figure}[h!]
		\centering 
		\includegraphics[width=\linewidth]{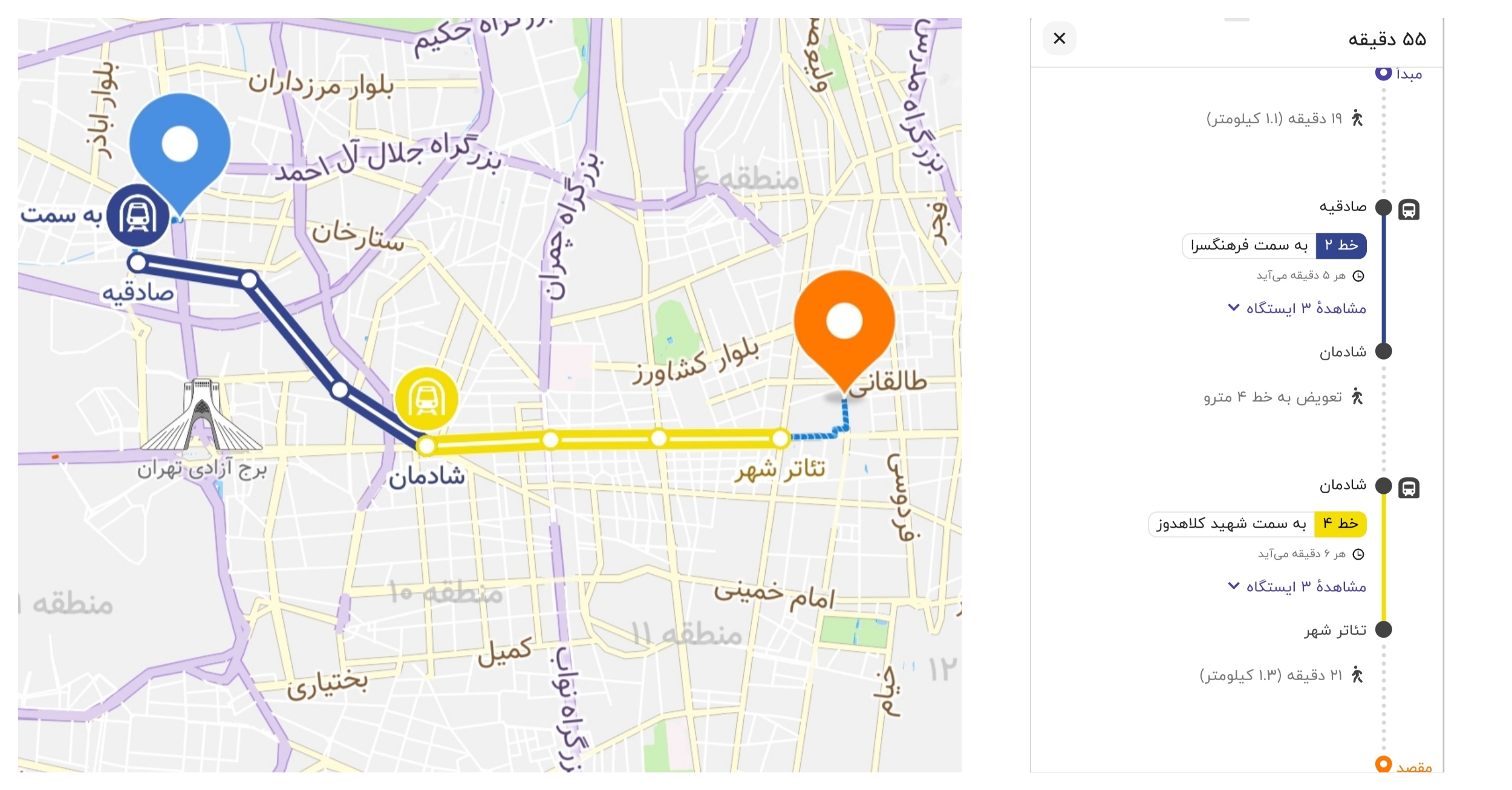}
		\caption{Fourth route in "Balad" application between Sadeghiyeh Square and Amir Kabir University}
		\label{fig:fig2_10}
\end{figure}

Figure \eqref{fig:fig2_10} shows the individual under consideration initially walked for 1100 meters. Then, they covered three subway stations and, after transferring lines, passed through 3 subway stations. Finally, they walked for 1300 meters. Initially, the probability of contracting the disease in each part of the route was calculated. 

\begin{equation}
	f_{41}=1-e^{-0.025299\times\frac{1100}{5000}}=0.0056
\end{equation}
\begin{equation}
	f_{42}=1-e^{-0.040155\times\frac{3\times3}{60}}=0.0060
\end{equation}
\begin{equation}
	f_{43}=1-e^{-0.040155\times\frac{3\times3}{60}}=0.0060
\end{equation}
\begin{equation}
	f_{44}=1-e^{-0.025299\times\frac{1300}{5000}}=0.0066
\end{equation}

Finally, using the equation \eqref{eq0}, the ultimate probability of contracting the disease for the entire route is determined. 

\begin{equation}
	f_4=1-(1-f_{41})(1-f_{42})(1-f_{43})(1-f_{44})=0.0239
\end{equation}

As a result, the probability of contracting the disease on the fourth route is 2.39

\subsubsection*{ Fifth Car Route}
\begin{figure}[h!]
		\centering 
		\includegraphics[width=\linewidth]{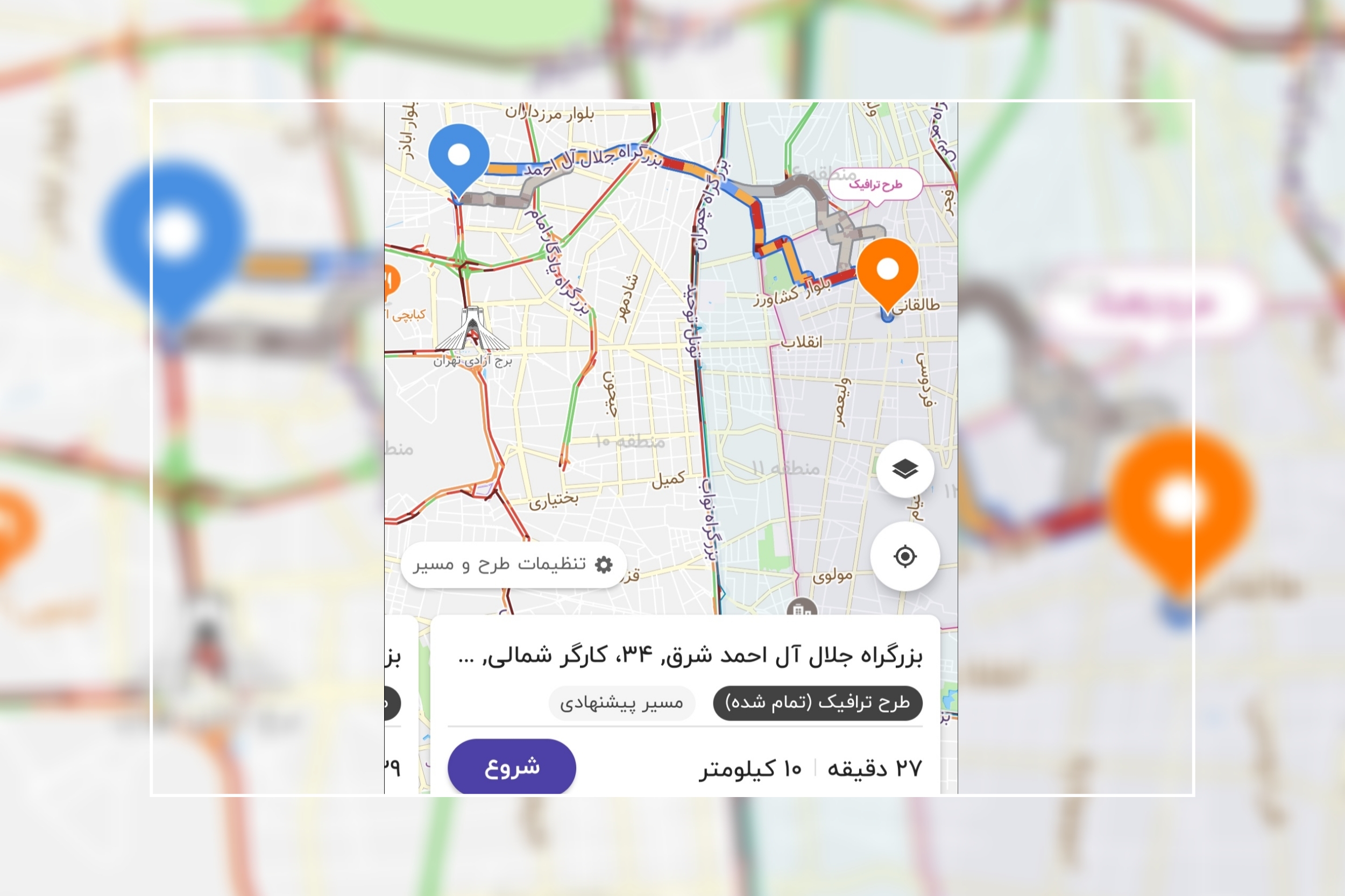}
		\caption{Fifth route in "Balad" application between Sadeghiyeh Square and Amir Kabir University}
		\label{fig:fig2_11}
\end{figure}

Figure \eqref{fig:fig2_11} shows that the individual under consideration has been in the car for 27 minutes. This calculation calculates the probability of contracting the disease as follows.

\begin{equation}
	f_5=1-e^{-0.407105\times\frac{27}{60}}=0.1674
\end{equation}
Calculating the probability of contracting the disease for the fifth route, we obtain 16.74

Among the proposed routes in the "Balad" application, the fourth route, a combination of pedestrian and subway routes, has the lowest probability of contracting the disease, with a probability of 0.5958. Therefore, this route is considered the safest during the pandemic. 

\section*{Summary, Conclusions and Suggestions}
In 2020 and with the peak of the Corona epidemic worldwide, it became clear that despite the progress that modern humans have made in various fields, they are still not ready to face a global pandemic and this disease affects various aspects of daily life. One of the aspects that was affected was the transportation issue. Considering the high probability of catching the disease during an epidemic, knowing the route that has the least chance of catching the disease can help people to reduce the chance of getting infected. In this article, an algorithm was tried to find the best route between an origin and a desired destination during the epidemic of diseases and specifically, Covid-19. The best path means the path that has the least chance of getting the disease. In the following, the obtained results are presented and summarized, and suggestions for future work are given.

\subsection*{Summary and Conclusion}
As mentioned earlier, this article was to find a route between a given origin and destination with the lowest disease contraction probability. To achieve this, we assumed that the final route consists of segments of different types, including "Walking", "Subway", "BRT (Bus Rapid Transit)", "City Bus" and "Car". In the second chapter, we focused on finding a mathematical relationship for the probability of disease contraction on each of these routes.
We derived a relationship for disease contraction probability on a route combining multiple segments. According to this relationship, if a route n consists of n-1 segments, each with a probability of disease contraction on a route i represented by Pi, the overall probability of disease contraction on the route n can be calculated as follows:
\begin{equation}
	P_n = 1-\prod_{i=1}^{n-1} (1- P_i)
	\label{eq4_1}
\end{equation}

According to the equation \eqref{eq4_1}, by considering the probability of disease contraction on sub-routes, it is possible to calculate the probability of disease contraction on a route composed of a combination of these sub-routes. Subsequently, we found a mathematical relationship for the probability of disease contraction on each route. For this purpose, we assumed that the probability of disease contraction depends on environmental density (x), the likelihood of a person being a carrier (y), and the duration the person spends in the environment (z). In this manner, we arrived at the equation \eqref{eq4_2} for disease contraction probability.
\begin{equation}
	f(x,y,z)=1-e^{c(x,y)z}
	\label{eq4_2}
\end{equation}

We calculated this probability by modeling the disease rate on an m-dimensional page where n infected individuals are present. This modeling assumes that a person without the disease passes through the page and stays in cell j for Tj seconds. At the end of the route, we calculate the probability of contracting the disease for this person, which is given by the equation \eqref{eq4_3}.
\begin{equation}
		h(k)= 1-\prod_{j=1}^m {1-g(\sum_{i=1}^n \frac{k}{r_{ji}^2})T_j}
		\label{eq4_3}
\end{equation}

In this equation, $r_{ij}$ represents the distance from the non-infected person in cell j to the infected person i. Additionally, the coefficient k in this formula depends on the route type and the individual's activity level. By substituting the equations \eqref{eq4_2} and \eqref{eq4_3} into the formula, we arrived at the following equation for the probability of contracting the disease.
\begin{equation}
	f(x,y,z)=1-e^{\sum_{j=1}^m{\sum_{i=1}^n{\frac{kz}{mr_{ij}^2}}}}
	\label{eq4_4}
\end{equation}

Finally, we utilized this equation to calculate the probability of contracting the disease for each sub-route. To calculate the coefficient k in equation \eqref{eq4_4} for the probability of contracting the disease in different environments and for different activity levels, we substituted the values in \cite{covid-19}. Since the k coefficient depends on the activity level, we used the k(E) coefficient in the formula. Ultimately, the probability of contracting the disease for each sub-route was obtained based on the formulas in Table \eqref{tab4_1}.

\begin{table} [!h]
  \centering
      \caption{Mathematical function of the probability of contracting the disease in different environments}
    \label{tab4_1}
    \begin{tabular}{|c|c|}
    \hline
         \textbf{environment} & \textbf{$f(m,n,z,E)$} \\
    \hline  
        pedestrian &
          $1-e^{\sum_{j=1}^m{\sum_{i=1}^n{\frac{(-0.00143853E+0.71455401)z}{mr_{ij}^2}}}}$  \\
    \hline 
         subway &
          $1-e^{\sum_{j=1}^m{\sum_{i=1}^n{\frac{(-0.00180107E+0.82402941)z}{mr_{ij}^2}}}}$  \\
    \hline 
        BRT &
          $1-e^{\sum_{j=1}^m{\sum_{i=1}^n{\frac{(-0.00149112E+0.38875338)z}{mr_{ij}^2}}}}$  \\
    \hline 
         city bus &
          $1-e^{\sum_{j=1}^m{\sum_{i=1}^n{\frac{(-0.00220276E+0.56565911)z}{mr_{ij}^2}}}}$  \\
    \hline 
       car &
          $1-e^{\sum_{j=1}^m{\sum_{i=1}^n{\frac{-2.729480z}{mr_{ij}^2}}}}$  \\
    \hline
    \end{tabular}
\end{table}

In the following, we first discussed the routing algorithm for selecting the most efficient route between an origin and a desired destination in a general context. The steps of this algorithm are as follows:
\begin{enumerate}
  \item 
  For each segment of the route, we calculate the values of parameters n (reliant on the hourly time of day under consideration), rmean (based on the length and width of the environment), E, and z.
  \item
  The probability of contracting the disease in each segment of the route, whether it is "Pedestrian," "Subway," "BRT," or "City Bus," is calculated using equation \eqref{eq4_5}. If it is "Car," the calculation is based on equation \eqref{eq4_6}.
  \begin{equation}
	1-e^{\frac{(k_1E+k_2)nz}{r_{mean}^2}}
	\label{eq4_5}
\end{equation}	

\begin{table} [!h]
  \centering
      \caption{$k_1$ and $k_2$ coefficients to calculate the probability of contracting a disease on pedestrian, subway, BRT and city buses}
    \label{tab4_2}
    \begin{tabular}{|c|c|c|}
    \hline
         \textbf{environment} & \textbf{$k_1$} & \textbf{$k_2$} \\
    \hline  
         pedestrian & $-0.00143853$ & $0.71455401$  \\
    \hline 
         subway & $-0.00180107$ & $0.82402941$  \\
    \hline 
        BRT & $-0.00149112$ & $0.38875338$  \\
    \hline 
         city bus & $-0.00220276$ & $0.56565911$  \\
    \hline
    \end{tabular}
\end{table}

\begin{equation}
	1-e^{\frac{-2.729480nz}{r_{mean}^2}}
	\label{eq4_6}
\end{equation}
  \item  
  By substituting the obtained probabilities in the previous step into the equation \eqref{eq4_1}, we calculate the probability of contracting the disease for the entire route.
  \item
  We repeat the above steps for the other recommended routes.
  \item
  The route with the lowest probability of contracting the disease is suggested as the preferred route. 
\end{enumerate}

In this algorithm, to calculate the number of infected individuals in each environment, we multiplied the probability of an individual being infected in the community by the total number of people in that environment. For example, the probability of an individual being infected with COVID-19 in Iran during the pandemic's peak was 0.8656
After determining the routing algorithm, we implemented this algorithm on the suggested routes in two routing applications, "Neshan" and "Balad." Consequently, the probability of contracting the disease for the recommended routes between "Sadeghiyeh Square" and "Amirkabir University" at 6:00 PM on a weekday was calculated. Six suggested routes were evaluated for the "Neshan" application, and the best route, consisting of pedestrian, subway, and BRT routes with a 2.62

\subsection*{Suggestions}

In this article, a routing algorithm was presented to choose the best route during the epidemic of diseases, including Covid-19. This algorithm is based on a mathematical model in which the probability of getting a disease can be calculated in any of the "Pedestrian", "Subway", "BRT", "City Bus" or "Car" routes. In this study, the coefficients of the mathematical model for the covid-19 disease were calculated, but this model can be extended to other infectious diseases that are transmitted through the air, and the coefficients of the model will be different according to the intensity and rate of prevalence.

In the routing algorithm, the probability of infection in the proposed routes between the origin and the specific destination given by the routing applications, are calculated. In this article, only the routes suggested in "Nashan" and "Balad" applications have been reviewed. For future work, this review can be done in other routing applications. Also, to make it practical, you can write a program that implements this algorithm on routing applications and suggests safe routes to users.

It is also possible to turn the problem into a "vehicle routing problem" that determines the best route with the least probability of getting the disease by having the origin and destination and without the need for backup routes in routing applications. To calculate the average distance between people in the environment, it was assumed that the environment is rectangular. For the "Pedestrian" environment, we considered the dimensions of the rectangle to be 4 x 20 meters, while according to the figure \eqref{fig:fig2_0} If there is a lot of congestion, this amount will no longer work. To solve this defect, in future works, instead of the average distance, the average of the probability function can be considered. Another thing that can be considered in the future is to consider the waiting time at the stations as well as the time to change subway lines. By adding this time into the calculations, the suggested routes may also change.

\bibliography{AmirBayat_BestRoute}

\bibliographystyle{Science}

\clearpage

\end{document}